\documentclass[twoside]{article}
\usepackage{amsmath,amssymb,times}
\usepackage{graphicx}
\usepackage[T1]{fontenc}
\usepackage[french]{babel}
\usepackage{amsthm}
\usepackage{amsfonts,bbm}
\usepackage{amssymb, amscd}
\usepackage{epsfig,color}
\usepackage{fancyhdr}
\def\shorttitle{\sc Gravitation quantique }
\leftmark
\def\shortauthor{\sc\leftmark}

\numberwithin{equation}{section}

\newcommand{\eps}{\varepsilon}

\newcommand{\HH}{\mathcal{H}}

\input xy
\xyoption{all}

\input epsf

\usepackage{amsmath}
\usepackage{amsfonts}
\usepackage{amssymb}
\usepackage{graphicx}
\newcommand{\lightline}{\rule{\textwidth}{0.2pt}}
\newenvironment{computation}[1]{\noindent \lightline\\\noindent\begin{small}\textit{#1}.}{\end{small}\noindent\lightline}

\newcommand{\Ref}[1]{(\ref{#1})}

\newcommand{\be}{\begin{equation}}
\newcommand{\ee}{\end{equation}}
\newcommand{\bea}{\begin{eqnarray}}
\newcommand{\eea}{\end{eqnarray}}
\def\d{\delta}
\def\f{\frac}
\def\om{\omega}
\def\w{\wedge}

\newcommand{\p}{\partial}
\newcommand{\n}{\nabla}

\begin{document}
\date{}
\title{Introductory lectures to loop quantum gravity\footnote{Lectures given at the 3eme Ecole de Physique Theorique de Jijel, Algeria, 26 Sep -- 3 Oct, 2009}}%

\author{\textbf{Pietro Don\'a$^{a,b}$ and Simone Speziale$^a$}
\smallskip \\ 
{\small $^a$\emph{Centre de Physique Th\'eorique\footnote{Unit\'e Mixte de Recherche (UMR 6207) du CNRS et des Universites Aix-Marseille I, Aix-Marseille II et du Sud Toulon-Var. Laboratoire affili\'e \`a la FRUMAM (FR 2291).}, CNRS-Luminy Case 907, 13288 Marseille Cedex 09, France}} \\
{\small $^b$  \emph{Scuola Normale Superiore, Piazza dei Cavalieri 7, 56126 Pisa, Italy}} 
}

\maketitle
\begin{abstract}

Nous pr\'esentons une introduction de base \`a la gravit\'e quantique \`a boucles, des variables ADM aux etats du r\'eseau de spin. Nous incluons une discussion sur la g\'eom\'etrie quantique sur un graphique fixe et sa relation avec une approximation discr\`ete de la relativit\'e g\'en\'erale.

\end{abstract}

\begin{center}
{\textbf{Abstract}} \vspace{0.5cm}

\begin{minipage}{10cm}
We give a standard introduction to loop quantum gravity, from the ADM variables to spin network states. We include a discussion on quantum geometry on a fixed graph and its relation to a discrete approximation of general relativity. 
\end{minipage}

\end{center}
\newpage
$ $
\newpage

 \tableofcontents

  \newpage

\section{A path to loop quantum gravity}

In order to motivate the loop approach, it is useful to begin 
with a review of the two main paths  to the quantization of general relativity: the covariant, or functional integral, approach, and the canonical, or Hamiltonian, approach.
Their partial successes and difficulties will shed light on the problem of quantum gravity.

\subsection{Covariant approach}\label{SecCov}

The basic goal of the covariant approach is to define a functional integral for General Relativity,
\be
\int {\cal D}g_{\mu\nu} e^{-i S_{EH}(g)},
\ee
where 
\begin{equation}
\label{EHaction}
S_{EH}(g_{\mu \nu}) = \frac{1}{16 \pi G} \int \mathrm{d}^4 x \, \sqrt{-g}\,  g^{\mu \nu}R_{\mu \nu}\left(\Gamma\left(g\right)\right)
\end{equation}
is the Einstein-Hilbert action. Here $g$ is the determinant of the metric, and $\Gamma(g)$ is the Levi-Civita connection entering the covariant derivative of vectors and tensors,
\be
\n_\mu v^\nu = \p_\mu v^\nu + \Gamma^\nu_{\rho\mu}(g) v^\rho, \qquad 
\Gamma^\nu_{\rho\mu}(g) = \f12 g^{\nu\lambda} [\p_\rho g_{\lambda\mu} + \p_\mu g_{\lambda\rho} - \p_\lambda g_{\rho\mu}].
\ee

From courses in QFT, we know how to properly define only the Gaussian integral
\begin{equation}\label{Gint}
\int \mathcal D \varphi \exp\left(-\int \frac{1}{2} \varphi \square \varphi \right).
\end{equation}
This corresponds to a free theory. An interacting theory can be treated in perturbation theory from a generating functional based on \Ref{Gint}. However, there is no quadratic term in the Einstein-Hilbert action (in fact, it is not even polynomial).
This obstruction can be bypassed if one performs a perturbative expansion of the metric. Consider the field redefinition 
\begin{equation}\label{fred}
g_{\mu \nu} = \eta_{\mu \nu} + h_{\mu \nu},
\end{equation}
where $\eta_{\mu \nu}$ is fixed to be the Minkowski metric, and $h_{\mu\nu}$ the new dynamical field. The introduction by hand of the background field $\eta_{\mu\nu}$ is crucial: if we treat $h_{\mu \nu}$ as a small fluctuation, a perturbative expansion of the action gives (in the De Donder gauge)
\begin{equation}
S_{EH}(g_{\mu \nu}) = \frac{1}{32 \pi G} \int \mathrm{d}^4 x \, h_{\mu \nu}\square h^{\mu \nu} + o(h^3).
\end{equation}
The background metric can be also used to define the Wick rotation. 
This action is amenable to the machinery of QFT. Among the main results of this background-dependent perturbative approach, let us highlight the following ones:
\begin{enumerate}
\item The quanta of the gravitational field are interpreted as massless spin-2 particles (the ``gravitons'').
\item In the low energy static limit, one can compute quantum corrections to the classical Newton potential \cite{Donoghue}, recovering the classical relativistic correction $o(G^2)$ and a new, purely quantum one $o(\hbar G^2)$:
\begin{equation}
V_N = - G \frac{m_1 m_2}{r}\left( 1+ G \frac{m_1 + m_2}{r} - \frac{127}{30 \pi^2} \frac{\hbar G}{r^2} + o(r^{-3})\right)
\end{equation}
\item As for other quantum field theories, one can define the S-matrix to describe the fundamental observables, and compute scattering amplitudes of gravitons with matter fields and with themselves \cite{Donoghue:1995cz}
\end{enumerate}

This brief list does not exhaust the interesting results obtained with this approach. On the other hand, there is a fundamental difficulty with it. As for any quantum field theory, infinities appear when considering the effects of arbitrarily small (``ultraviolet'') field fluctuations. These divergences are usually dealt with via the procedure of renormalization. This technique, successful for $\lambda \phi^4$ or gauge theories, fails in the case of gravity. This can be expected from dimensional arguments, and the rigorous proof that the perturbative quantization of general relativity fails because of non-renormalizable ultraviolet divergences was obtained  in the late eighties, by Goroff and Sagnotti \cite{GorSagn}.

This means that the theory constructed as sketched above can be used for low-energy calculations, but it would be inconsistent if taken seriously at all energy scales. Heuristically, the inconsistence can be explained as follows. The key field redefinition \Ref{fred} assumes that we can quantize the field $h_{\mu\nu}$, and that its dynamics takes place on a fixed classical background $\eta_{\mu\nu}$. 
However, as we increase the energy of the fluctuations its backreaction increases, and it becomes inconsistent to assume that the background stays fixed, unperturbed.
Although gravitons capture correctly the low-energy physics of the gravitational field, they might not be the right quantities to describe the quanta of gravity at Planckian energies.

\subsection{Canonical approach}
The canonical formalism is based on second quantization, and describes the quantum theory in terms of functionals of the fields, e.g. $\Psi[\phi]$ for a scalar field $\phi$, like the familiar function of the configuration variables $\psi[x]$ in quantum mechanics. The dynamics is described by the quantum Hamiltonian $\hat H$ and the Sch\"odinger equation 
\be
i\hbar \f{\p}{\p t} \Psi[\phi] = \hat H\left(\phi, \f{\d}{\d \phi}\right) \Psi[\phi]
\ee
For the reader less familiar with this approach, see \cite{Hatfield}.
Given the overwhelming success of the functional integral approach to the quantization of (non-gravitational) field theories, the canonical formalism is often underappreciated in many courses, and a certain taste of ``old fashion'' is associated with it.
It was applied to general relativity by Arnowitt, Deser  and Misner (ADM), Dirac, Wheeler and De Witt, among many others. The loop approach is related to this original idea, which we will now review in some details. For more on it, see e.g. \cite{Gravitation}.

\subsubsection{ADM formalism}

In order to put the Einstein-Hilbert action into canonical form one needs to identify the variables which are canonically conjugated, and then perform the Legendre transform. To that end, we make the assumption that $\mathcal M$ has the topology $\mathcal M \cong \mathbb R \times \Sigma$ where $\Sigma$ is a fixed three-dimensional manifold of arbitrary topology and spacelike signature. 
This poses no restrictions if we require that $\mathcal M$ has no causally disconnected region (more precisely that $\mathcal M$ is globally hyperbolic). In fact by a theorem due to Geroch and improved by Bernal and Sanchez \cite{Geroch,BerSan} if the space-time is globally hyperbolic then it is necessarily of this kind of topology.

Having made this assumption, one knows that $\mathcal M$ foliates into a one-parameter family of hypersurfaces $\Sigma_t = X_t(\Sigma)$ embeddings of $\Sigma$ in $\mathcal M$.
The foliation allows us to identify the coordinate $t\in \mathbb R$ as a time parameter. Notice however that this ``time'' should not be regarded as an absolute quantity, because of the diffeomorphism invariance of the action. A diffeomorphism $\phi \in \mathrm{Diff}\left( \mathcal M \right)$ maps a foliation $X$ into a new one $X' = X \circ \phi$, with a new time parameter $t'$. Conversely, we can write a general diffeomorphism $\phi \in \mathrm{Diff}\left( \mathcal M \right)$ as composition of different foliations, $\phi = X' \circ X^{-1}$. 
Hence, we can always work with a chosen foliation, but the diffeomorphism invariance of the theory will guarantee that physical quantities are independent of this choice. A theory of spacetime which unlike general relativity has a preferred foliation and thus a preferred time, is a theory which breaks diffeomorphism invariance.

Given a foliation $X_t$ and adapted ADM coordinates $(t,x)$, we can define the time flow vector
\begin{equation}
\tau^\mu(x) \equiv \frac{\partial X_t^\mu (x)}{\partial t} = (1,0,0,0).
\end{equation}
This vector should not be confused with the \emph{unit normal} vector to $\Sigma$, which we denote $n^\mu$. They are both timelike, $g_{\mu\nu}\tau^\mu\tau^\nu=g_{00}$ and $g_{\mu\nu}n^\mu n^\nu=-1$, but they are not parallel in general. Let us decompose $\tau^\mu$ into its normal and tangential parts, 
\begin{equation}
\label{timeflow}
\tau^\mu(x) = N(x)n^\mu(x) + N^\mu(x).
\end{equation}
\begin{center}
\includegraphics[width=6cm]{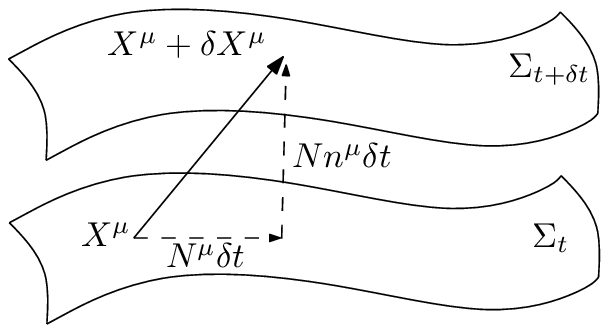}\\
\end{center}
It is convenient to parametrize $n^\mu=(1/N, -N^a/N)$, so that $N^\mu=(0,N^a)$.
$N$ is called lapse function, and $N^a$ shift vector.
In terms of lapse and shift, we have
\begin{align*}
&g_{\mu\nu}\tau^\mu\tau^\nu = g_{00} = -N^2 + g_{ab} N^a N^b, \\
&g_{\mu\nu} \tau^\mu N^\nu = g_{0b} N^b = g_{\mu\nu}(Nn^\mu + N^\mu) = g_{ab} N^a N^b 
\ \longrightarrow \ g_{0a} = g_{ab}N^b \equiv N_a.
\end{align*}
Using these results, the metric tensor can be written as
\begin{equation}\label{gADM}
\mathrm d s^2 = g_{\mu \nu} \mathrm dx^\mu \mathrm dx^\nu = - \left( N^2-N_a N^a\right) \mathrm d t^2 + 2 N_a \mathrm d t \mathrm d x^a + g_{ab}  \mathrm dx^a \mathrm dx^b,
\end{equation}
where $a=1,2,3$ are spatial indices and are contracted with the 3-dimensional metric $g_{ab}$.

Notice that the spatial part $g_{ab}$ is \emph{not} in general the intrinsic metric on $\Sigma_t$. The latter is given by 
\begin{equation}
q_{\mu \nu} = g_{\mu \nu} - n_\mu n_\nu.
\end{equation}
Tensors on the spatial slide $\Sigma$, since their scalar product with $n$ vanishes, can be equivalently contracted with $g$ or $q$. 
The quantity $q^\mu_\nu = g^{\mu\rho} q_{\rho\nu}$ acts as a projector on $\Sigma_t$, allowing us to define the tensorial calculus on $\Sigma_t$ from the one on $\mathcal M$. 
An important quantity is the extrinsic curvature of $\Sigma_t$,
\begin{equation}
K_{\mu \nu} = q^{\mu'}_\mu q^{\nu'}_\nu \nabla_{\mu'}n_{\nu'}.
\end{equation}
This tensor is symmetric and it is connected to the Lie derivative of the intrinsic metric $\mathcal L_n q_{\mu \nu}=2K_{\mu\nu}$.
It enters the relation between the Riemann tensor of $\Sigma_t$ ($\mathcal R$) and that of $\mathcal M$ ($R$),
\begin{equation}
\label{GaussCodazzi}
\mathcal R ^\mu_{\nu\rho\sigma} = q_{\mu'}^\mu q^{\nu'}_\nu q^{\rho'}_\rho q^{\sigma'}_\sigma R^{\mu'}_{\nu'\rho'\sigma'} - K_{\nu\sigma}K^\mu_\rho - K_{\nu\rho}K^\mu_\sigma
\end{equation}
This formula, proved in the Appendix and known as the Gauss-Codazzi equation, lets us rewrite the action (\ref{EHaction}) in the following form,\footnote{We are assuming for simplicity of exposition that $\Sigma$ has no boundary, the case with boundary is treated extensively in \cite{HaHo} and \cite{HaHu}.}
\begin{equation}
S = \int \mathrm dt \int_{\Sigma} \mathrm d^3x \sqrt{q}N\left[\mathcal R - K^2 + \mathrm{Tr}(KK) \right].
\end{equation}

A key consequence of this analysis is the fact that time derivatives of $N$ and $N^a$ do not enter in the Lagrangian $L$. This implies that $N$ and $N^a$ are Lagrange multipliers, with null conjugate momenta ${\d L}/\d \dot N={\d L}/\d \dot N^a=0$. The true dynamical variables are the spatial components $q_{ab}$ only, with conjugate momenta 
\begin{equation}
\pi ^{ab} \equiv \frac{\delta L}{\delta \dot q_{ab}} = \sqrt{q}\left( K^{ab} - K q^{ab}\right).
\end{equation}
One can finally evaluate the Legendre transform, which after some calculations gives 
\begin{equation}
\label{EHaction2}
S_{EH}\left(q_{ab},\pi^{ab},N,N^a\right)=\frac{1}{16 \pi G}\int \mathrm d t \int \mathrm d ^3 x\; 
\left[\pi^{ab}\dot{q}_{ab} -  N^a H_a - N H \right],
\end{equation}
where (the covariant derivative $\n$ contains only spatial indexes)
\be
 H_a = - 2 \sqrt{q}\, \n_b \left( \frac{\pi^b_a}{\sqrt{q}}\right)
\end{equation}
and 
\begin{equation}\label{ADMHam}
 H = \frac{1}{\sqrt{q}} \, G_{abcd} \pi^{ab}\pi^{cd} - \sqrt{q}\, \mathcal R, \qquad
G_{abcd} = q_{ac}q_{bd}+q_{ad}q_{bc}- q_{ab}q_{cd}.
\end{equation}
The quantity $G_{abcd}$ is often called the supermetric, or DeWitt metric.
The variation of the action \eqref{EHaction2} with respect to the Lagrange multipliers gives the equations
\begin{equation}
\label{constraints}
 H_a(q,\pi)=0, \qquad \qquad  H(q,\pi) = 0,
\end{equation}
which are called respectively vector, or space-diffeomorphism constraint, and  scalar or Hamiltonian constraint. In the following we will also use the compact notation $ H_\mu=( H, H_a)$. Physical configurations, also called on-shell configurations with particle physics jargon, must satisfy these constraints.

From \Ref{EHaction2} we see that the Hamiltonian of general relativity is
\begin{equation}
{\bf H} =\frac{1}{16 \pi G} \int \mathrm d ^3 x\;  N^a H_a + N H.
\end{equation}
This Hamiltonian is peculiar, since it is proportional to the Lagrange multipliers and thus \emph{vanishes} on-shell.
Hence, there is no dynamics and no physical evolution in the time $t$. This puzzling absence of a physical Hamiltonian is in fact a consequence of what we discussed earlier: the diffeomorphism-invariance of the theory tells us that $t$ is a mere parameter devoid of an absolute physical meaning, thus there is no physical dynamics in $t$.
This is the root of the \emph{problem of time} in general relativity. For discussions of this problem, see \cite{Isham} and \cite{RovelliBook}.

\subsubsection{Symplectic structure}

From now on, we work in units $16 \pi G  = 1$.
The Hamiltonian formulation \Ref{EHaction2} allows us to study the phase space of general relativity. It is parametrized by the pair $(q_{ab}, \pi^{ab})$, with canonical Poisson brackets
\begin{equation}\label{piq}
\left\lbrace \pi^{ab} (t,x), q_{cd} (t,x')\right\rbrace = \delta^a_{(c} \delta^b_{d)} \delta(x-x').
\end{equation}
From there we can evaluate the following brackets among the constraints (for the explicit calculation see in the Appendix)
\begin{align}\label{ConstraintAlgebra}
&\left\lbrace  H_a(x),  H_b(y)\right\rbrace =  H_a (y)\partial _b \delta(x-y) -  H_b(x)\partial '_a\delta(x-y)\nonumber \\
&\left\lbrace  H_a(x),  H(y)\right\rbrace  =  H(x) \partial_a \delta(x-y) \\
&\left\lbrace  H(x),  H(y)\right\rbrace =  H^a (y)\partial _a \delta(x-y) -  H^a(y)\partial '_a\delta(x-y)\nonumber\end{align}
Notice that the right-hand sides vanish on the constraint surface \Ref{constraints}. This means that the Poisson flows generated by the constraints preserve the constraint hypersurface. Constraints with this characteristic are said to be \emph{first class}, as opposed to \emph{second class} constraints whose Poisson brackets do not vanish on-shell. First class constraints generate gauge transformations on the constraint surface (See e.g. \cite{Mukhanov:1994zn} for details). 

To see what the gauge transformations look like in our case, consider the smearing of the constraints
\begin{equation}
 H(\vec N) = \int_\Sigma  H^a(x) N_a(x) \mathrm d ^3 x, \hspace*{1cm} 
 H(N) = \int_\Sigma  H(x) N(x) \mathrm d ^3 x.
\end{equation}
An explicit computation (see Appendix) shows that
\begin{equation}
\left\lbrace  H(\vec N),q_{ab}\right\rbrace = \mathcal L_{\vec{N}}q_{ab}, 
\hspace{1cm} \left\lbrace  H(\vec N),\pi^{ab}\right\rbrace = \mathcal L_{\vec{N}}\pi^{ab}
\end{equation}
which means that  the vector constraint is the generator of space-diffeomorphism on $\Sigma$. The situation is somewhat subtler for the Hamiltonian constraint. We now have  (see Appendix)
\bea
\left\lbrace  H(N),q_{ab}\right\rbrace &=& \mathcal L_{\vec{n}N}q_{ab}, \\ 
\left\lbrace H(N),\pi^{ab}\right\rbrace &=& \mathcal L_{\vec{n}N}\pi^{ab} + \frac12{q^{ab}N H} - 2N\sqrt{q} q^{c[a}q^{b]d}R_{cd}
\eea
The first bracket is the action of time diffeomorphisms on $q_{ab}$. The second bracket gives the action of time diffeomorphisms on $\pi_{ab}$, but contains also two extra pieces. These vanish if $H=0$ and $R_{cd}=0$, namely on the constraint surface and for physical solutions (recall that in vacuum Einstein's equations read $R_{\mu\nu}=0$).
Therefore, we conclude that  the constraints $H^\mu$ are the generators of the spacetime diffeomorphism group $\mathrm{Diff}(\mathcal{M})$ on physical configurations.

For general configurations, \Ref{ConstraintAlgebra} defines the algebra of hypersurface deformations, often called Dirac algebra or Bargmann-Komar algebra. A characteristic of this algebra is that it is not a Lie algebra. In fact, let us look at the smeared Poisson bracket
\begin{equation}
\left\lbrace H(N_1), H(N_2)\right\rbrace = H \left(g^{ab}(N_1 \partial_b N_2 - N_2 \partial_b N_1)\right).
\end{equation}
We see that outside the constraint surface, the "structure constants" on the right-hand side contain the field $g^{ab}$ itself. Hence they are not constants at all, and the algebra \Ref{ConstraintAlgebra} is not a Lie algebra, unlike $\mathrm{Diff}(\mathcal{M})$.  
Instead, \Ref{ConstraintAlgebra} shows that when we introduce a foliation to define the canonical formalism, 
we still have the symmetry of diffeomorphisms, which acts changing the foliation, and this new one, which acts deforming it. The two symmetries coincide on physical configurations.

\subsubsection{Constraints and physical degrees of freedom}

A simple counting of the number of degrees of freedom of general relativity can be done in the covariant perturbative approach described in Section \ref{SecCov}. The structure of the linearized field equations shows that only two components of $h_{\mu\nu}$ propagate, which correspond to the two helicities of a massless spin 2 particle. 

An advantage of the canonical formalism is that it allows us to confirm this counting in a more general and robust way. Recall in fact that in classical physics each point in phase space (i.e. ``initial position and momentum'') characterizes a physical trajectory, and  the number of degrees of freedom is defined to be half the dimensionality of the phase space. In constrained theories, such as general relativity but also gauge theories, one has to be careful with the constraints. To that end, it is customary to distinguish a notion of \emph{kinematical} phase space, and \emph{physical} phase space. The kinematical phase space is the one defined by the Poisson structure of the theory. In our case, the space $\left( q_{ab},\pi^{ab}\right)$, with Poisson brackets \Ref{piq}.
The dimension of this space is $(6+6)\cdot \infty^3 = 12 \cdot \infty^3$.

On this space, the constraints define a hypersurface where they are satisfied, i.e. the space of $\left( q_{ab},\pi^{ab}\right)$ such that $H^\mu(q,\pi)\equiv0$. We call this the constraint surface. Its dimensionality is $(12-4)\cdot \infty^3 = 8 \cdot \infty^3$. The fact that the algebra of constraints is first class guarantees that the gauge transformations generated by the constraints preserve the constraint surface. We shall refer to the trajectories drawn by the gauge transformations as orbits. Points along one orbit correspond to the \emph{same} physical configuration, only described in different coordinate systems. Hence, to select the physical degrees of freedom we have to divide by the gauge orbits, in a manner identical to what happens in gauge theories. Since the orbits span a dimension 4 manifold at each space point, dividing by the orbits gives $(8-4)\cdot \infty^3 = 4 \cdot \infty^3$. This is the physical phase space. 
It has four dimensions per space point, namely the theory has two physical degrees of freedom per space point (or simply two degrees of freedom, for brevity, with the space dependence tacitly implied), a result consistent with the linearized analysis. 
See e.g. \cite{Mukhanov:1994zn} or \cite{HT} for more details.

\begin{center}
\includegraphics[width=10cm]{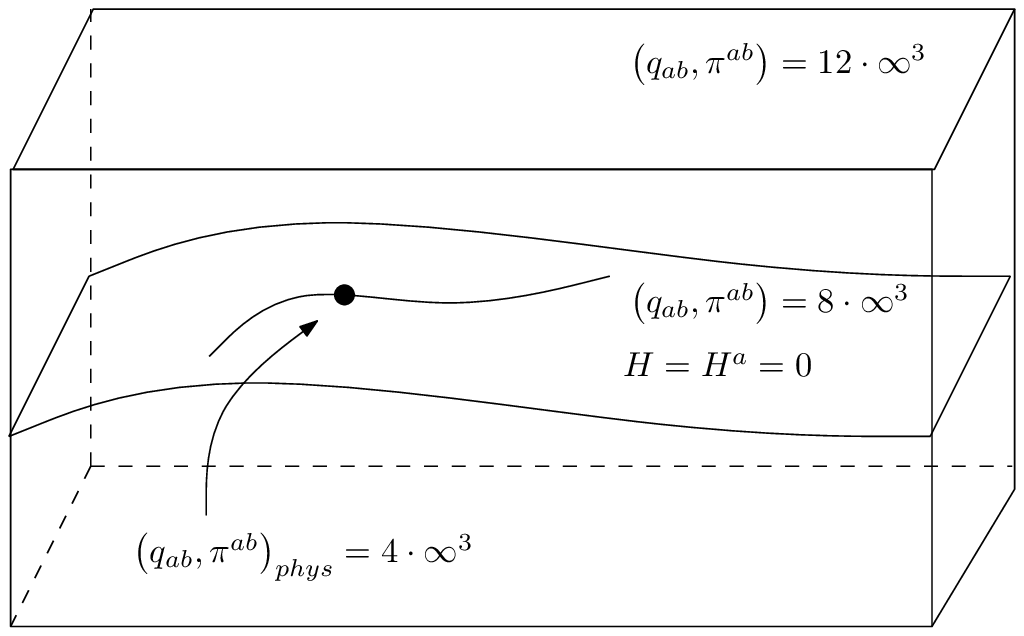}\\
\end{center}

This has far as the counting goes. However, in the case of the linearized analysis we are also able to identify the 2 degrees of freedom as the two helicities, and associate a physical trajectory to each point in phase space, thanks to the fact that we are able to solve the dynamics. Compare this to the mechanics of a point particle: the dynamics is given by $\phi_{[\phi_0,p_0]}(t)$, and the degree of freedom corresponds to varying the initial conditions $(\phi_0,p_0)$, spanning in this way all possible physical evolutions.

Therefore, if we want to know what the two physical degrees of freedom of general relativity are, we need to control the general solution of the theory. This is a formidable task due to the high non-linearity of the equations, and in spite of the effort in this direction, still little is known. See \cite{Isham2} for a review of some attempts.

There is a particular reformulation of the dynamics that plays a role in quantization.
Let us go back to the Gauss-Codazzi equation (\ref{GaussCodazzi}). 
A simple manipulation gives the so-called first Gauss-Codazzi identity,
\begin{equation}
\label{1GCidentity}
\mathcal R + K^2 - \mathrm{Tr}(KK) = 2 n^\mu n^\nu G_{\mu\nu},
\end{equation}
where $G$ is the Einstein tensor. One can show (but we will not do it here) that the left hand side of (\ref{1GCidentity}) is equivalent to a linear combination of the constraints. This means that if $g_{\mu \nu}$ satisfies the constraints $H=H^a=0$ on \emph{any spacelike Cauchy hypersurface}, than it also satisfy all ten Einstein equations $G_{\mu \nu} = 0$. This is a crucial point: \emph{the whole dynamical content of general relativity is in those four constraints}.

\subsubsection{Dirac's quantization program}
The fact that the constraints include the whole dynamics is at the heart of the theory. It is heavily exploited by the approach to quantization proposed by Dirac, which is based on a definition of dynamical physical states as the ones annihilated by the constraints. The procedure can be schematically divided in three steps:
\begin{enumerate}
\item[(i)] Find a representation of the phase space variables of the theory as operators in an auxiliary ``kinematical'' Hilbert space $\mathcal H_{kin}$, satisfying the standard commutations relations
\begin{equation}
\left\lbrace \cdot, \cdot \right\rbrace \longrightarrow \frac{1}{i \hbar}\left[ \cdot, \cdot \right];
\end{equation}
\item[(ii)] Promote the constraints to operators $\hat H^\mu$ in $\mathcal H_{kin}$;
\item[(iii)] Characterize the space of solutions of the constraints $\mathcal H_{phys}$,
\be
\hat  H^\mu \psi = 0 \qquad \forall \psi \in  \HH_{	phys}.
\ee
\end{enumerate}
These steps should then be completed with an explicit knowledge of the scalar product in $ \HH_{phys}$ and a physical interpretation of the quantum observables.

Dirac's procedure is more general than gravity: it applies to any fully constrained system. Let us now try to apply it to the ADM formulation of general relativity.
We then look for a space of functionals carrying a representation of the quantum Poisson algebra
\begin{eqnarray}
\left[ \hat q_{ab}(x), \hat \pi^{cd}(y)\right] &=& i \hbar \delta^{cd}_{(ab)}\delta^3(x,y), \\
\nonumber \left[ \hat q_{ab}(x), \hat g_{cd}(y)\right] &=& 0, \\
\nonumber \left[ \hat \pi^{ab}(x), \hat \pi^{cd}(y)\right] &=& 0.
\end{eqnarray}
Formally, we can proceed by analogy with better known cases, such as a scalar field theory, and consider a Schr\"odinger representation
$ \hat q_{ab}(x) =  q_{ab}(x)$, $\hat \pi^{ab}(x) = - i \hbar \frac{\delta}{\delta  q_{ab}(x)}$, acting on wave functionals
\begin{equation}
\psi[q_{ab}(x)]
\end{equation}
of the 3-metric. This procedure, which works well for the scalar field \cite{Hatfield}, encounters a number of difficulties when applied to the gravitational context. For instance, to define the (kinematical) Hilbert space we need a scalar product, formally
\begin{equation}
\int \mathrm d g \, \overline{\psi[g]} \, \psi ' [g] \equiv \left\langle \psi \mid \psi'\right\rangle .
\end{equation}
However, there is no Lebesgue measure on the space of metrics modulo diffeomorphisms that we can use to define d$g$. Without this, we can not even check that $\hat q_{ab}(x) $ and $\hat \pi^{ab}(x)$ are hermitian, nor that $\hat q_{ab}(x) $ has positive definite spectrum, as needed to be a spacelike metric.

Let us ignore these and other similar issues, and try to proceed formally assuming that a well-defined $\HH_{kin}$ exist. The next step is to promote the constraints \Ref{constraints} to operators, and characterize their space of solutions. Let us proceed in two halves,
\be \HH_{kin} \ \ \xrightarrow{\displaystyle \hat H^a=0} \ \  \HH_{Diff} \ \ \xrightarrow{\displaystyle \hat H=0\phantom{^a} } \ \ \HH_{phys}.
\ee
Consider first the vector constraint. In the Schr\"odinger representation defined above, the smeared version gives
\begin{eqnarray}
\hat  H (N_a) \psi[q_{ab}] = 2 i \hbar \int_\Sigma \mathrm{d}^3 x \n_b N_a \frac{\delta \psi}{\delta q_{ab}}=0,
\end{eqnarray}
after an integration by parts.
This implies straightforwardly that $$\psi[q_{ab} + 2 \n_{(a}N_{b)}]\equiv \psi[q_{ab}],$$
namely the solution of the vector constraint are those functionals of the metric invariants under diffeomorphism. This is very nice, as it realizes at the quantum level the correct action of the classical constraints. However, the space of solutions $ \HH_{Diff}$ is again ill-defined, since it inherits from the kinematical one the lack of measure theory or other means of control over it. 
 
For the Hamiltonian constraint we can write
\begin{equation}\label{WdW}
\hat  H \psi[q_{ab}] = \left[ -\f{\hbar^2}{2} G^{abcd} :\f{1}{\sqrt{\mathrm{det}\hat g}} \f{\d^2}{\d q_{ab}(x) \d q_{cd}(x)}: - \sqrt{\mathrm{det}\hat g} R(\hat g)\right] \psi[q_{ab}],
\end{equation}
where the colon $:$ means that an ordering of the operators needs to be prescribed. The situation is more complicated for the Hamiltonian constraint, since it requires the definition of products of operators at the same point, notoriously very singular objects. 
The formal expression \Ref{WdW} is known as the Wheeler-DeWitt equation. 
Even if manage to give a suitable ordering prescription and regularize the differential operator, the problem with the equation is that we do not have any characterization of the solutions, not even formally as for the diffeo-constraints above. And of course, again no clue on the knowledge of the physical Hilbert space and scalar product (see however minisuperspace models \cite{Hartle}).

Loop quantum gravity is an approach to the problem which improves significantly the situation, and gives a number of answers to these open questions. The key to LQG and to such improvement is surprisingly simple: instead of changing the gravitational theory or the quantization paradigm, we just use different variables to describe gravity.
After all, we are familiar with the fact that not all choices of fundamentals variables work out as well when quantizing a classical theory.
Consider for instance the harmonic oscillator. Classically, the most elegant description of the system is in terms of action-angle variables, which parametrize the phase space as $\left\lbrace \phi, J\right\rbrace =1$ instead of $\left\lbrace q, p\right\rbrace =1$. However, it is more convenient to quantize the system using the $(q,p)$ variables 
than the action-angle ones, which require extra care in constructing the operator algebra and dealing with unitarity.

We now introduce the variables that allows us to reformulate general relativity in a way more amenable to Dirac's quantization procedure.


\subsection{Appendix }
In this Appendix we report some calculations which are used in the main text.

\begin{computation}{Proof of the Gauss-Codazzi equation}
We start the computation from the characterization of $\mathcal R$ in terms of covariant derivatives $\tilde \nabla$ defined as the projection of $\nabla$ on $\Sigma$ via $q^\mu_\nu$
\begin{equation}
\label{GCproof1}
2 \tilde \n_{[\mu} \tilde \n_{\nu]} w_\rho = - w_{\rho'}\mathcal R^{\rho'}_{\rho\mu\nu}
\end{equation}

\begin{equation}
 \tilde \n_{\mu} \tilde \n_{\nu} w_\rho= \tilde \n_{\mu}\left(q^{\nu'}_\nu q^{\rho'}_\rho \nabla_{\nu'} w_{\rho'}\right) =
q^{\mu'}_\mu q^{\nu''}_\nu q^{\rho''}_\rho\nabla_{\mu'}\left(q^{\nu'}_{\nu''} q^{\rho'}_{\rho''} \nabla_{\nu'} w_{\rho'}\right) 
\end{equation}
it's useful to notice that 
\begin{align}
q^{\mu'}_{\mu} q^{\nu'}_{\nu} \n_{\mu'}q^\rho_{\nu'} &= q^{\mu'}_{\mu} q^{\nu'}_{\nu} \n_{\mu'}\left( g^\rho_{\nu'} + n^\rho n_{\nu'}\right) =\\
&= q^{\mu'}_{\mu} q^{\nu'}_{\nu} \n_{\mu'}g^\rho_{\nu'} + q^{\mu'}_{\mu} q^{\nu'}_{\nu}  n_{\nu'} \n_{\mu'}n^\rho + q^{\mu'}_{\mu} q^{\nu'}_{\nu} n^\rho \n_{\mu'} n_{\nu'}= \\
&= n^\rho q^{\mu'}_{\mu} q^{\nu'}_{\nu} \n_{\mu'} n_{\nu'} = K_{\mu\nu}n^\rho
\end{align}
we used the definition of extrinsic curvature, orthogonality ($q^{\nu'}_{\nu}  n_{\nu'}=0$) and metric compatibility ($\nabla_{\mu'}g^\rho_{\nu'}=0$).
\begin{equation}
\tilde \n_{\mu} \tilde \n_{\nu} w_\rho= q^{\mu'}_\mu q^{\nu'}_\nu q^{\rho'}_\rho \nabla_{\mu'}\nabla_{\nu'} w_{\rho'} + q^{\rho'}_\rho n^{\nu'} K_{\mu\nu} \nabla_{\nu'}w_{\rho'} + q^{\nu'}_\nu K_{\mu\rho} n^{\rho'} \nabla_{\nu'} w_{\rho'}
\end{equation}
in particular if we consider $w\in \Sigma$ thus $n^{\rho'}w_{\rho'}=0$
\begin{equation}
n^{\rho'} \nabla_{\nu'} w_{\rho'} = \nabla_{\nu'} n^{\rho'}  w_{\rho'} -  w_{\rho'} \nabla_{\nu'} n^{\rho'} = -  w_{\rho'} \nabla_{\nu'} n^{\rho'}
\end{equation}
Antisymmetrizing in $a$,$b$
\begin{align}
\tilde\n_{[\mu} \tilde\n_{\nu]} w_\rho &= q^{\mu'}_{[\mu} q^{\nu'}_{\nu]} q^{\rho'}_\rho \nabla_{\mu'}\nabla_{\nu'} w_{\rho'} + q^{\rho'}_\rho n^{\nu'} K_{[\mu\nu]} \nabla_{\nu'}w_{\rho'} - q^{\nu'}_{[\nu} K_{\mu]\rho} w_{\rho'} \nabla_{\nu'}n^{\rho'} =\\
&= q^{\mu'}_{\mu} q^{\nu'}_{\nu} q^{\rho'}_\rho \nabla_{[\mu'}\nabla_{\nu']} w_{\rho'} -  K^{\rho'}_{[\nu} K_{\mu]\rho} w_{\rho'} = \\
&= - \frac{1}{2} q^{\mu'}_{\mu} q^{\nu'}_{\nu} q^{\rho'}_\rho R^{\rho''}_{\rho'\mu'\nu'}w_{\rho''} - K^{\rho'}_{[\nu} K_{\mu]\rho} w_{\rho'}
\end{align}
or in terms of the Riemann tensor using \eqref{GCproof1} 
\begin{equation}
\mathcal R^{\rho'}_{\rho\mu\nu}w_{\rho'} =\left(  q^{\mu'}_{\mu} q^{\nu'}_{\nu} q^{\rho''}_\rho R^{\rho'}_{\rho''\mu'\nu'} + 2K^{\rho'}_{[\nu} K_{\mu]\rho}\right)  w_{\rho'}
\end{equation}
this equation state for $w\in\Sigma$ for make it true for all $v \in \mathcal M$ it is sufficient to take $w_\mu = q^{\mu'}_\mu v_{\mu'}$
\begin{equation}\mathcal R^{\sigma}_{\rho\mu\nu} =  q^{\mu'}_{\mu} q^{\nu'}_{\nu} q^{\rho'}_\rho q_{\sigma'}^\sigma R^{\sigma'}_{\rho'\mu'\nu'}  + 2K^{\sigma}_{[\nu} K_{\mu]\rho}
\end{equation}
Or equivalently
\begin{equation}
\mathcal R_{\sigma\rho\mu\nu} =  q^{\mu'}_{\mu} q^{\nu'}_{\nu} q^{\rho'}_\rho q^{\sigma'}_\sigma R_{\sigma'\rho'\mu'\nu'}  + 2K_{\sigma[\nu} K_{\mu]\rho}
\end{equation}
\begin{align}
\mathcal R &= q^{\sigma\mu}q^{\nu\rho}q^{\mu'}_{\mu} q^{\nu'}_{\nu} q^{\rho'}_\rho q^{\sigma'}_\sigma R_{\sigma'\rho'\mu'\nu'}  + q^{\sigma\mu}q^{\nu\rho}K_{\sigma\nu} K_{\mu\rho} - q^{\sigma\mu}q^{\nu\rho}K_{\sigma\mu} K_{\nu\rho}=\\
&=q^{\sigma'\mu'}q^{\nu'\rho'} R_{\sigma'\rho'\mu'\nu'}  + \mathrm{Tr}(KK) - K^2=\\
&=R + 2 n^{\nu'}n^{\rho'}R_{\rho'\nu'} + n^{\sigma'}n^{\mu'}n^{\nu'}n^{\rho'}R_{\sigma'\rho'\mu'\nu'}  + \mathrm{Tr}(KK) - K^2
\end{align}

the term $n^{\sigma'}n^{\mu'}n^{\nu'}n^{\rho'}R_{\sigma'\rho'\mu'\nu'}$ vanishes because of the symmetries of the Riemann tensor, using the fact that $n^{\nu'}n^{\rho'}g_{\rho'\nu'} = -1$ we can write
\begin{equation}
\mathcal  R =2 n^{\nu'}n^{\rho'}\left( R_{\rho'\nu'} - \frac{g_{\rho'\nu'}}{2}R\right)  + \mathrm{Tr}(KK) - K^2
\end{equation}
That is the first Gauss-Codazzi equation.\\
\lightline\\
{\it Constraint Algebra. }
We define the Poisson bracket as 
\begin{equation}\left\lbrace A(y),B(z)\right\rbrace = \int \mathrm{d}^3 x \frac{\delta A(y)}{\delta q_{ab} (x)}\frac{\delta B(z)}{\delta \pi^{ab} (x)} - \frac{\delta B(z)}{\delta q_{ab} (x)}\frac{\delta A(y)}{\delta \pi^{ab} (x)}\end{equation}
We start the computation from the diffeomorphism constraint
\begin{equation}H(\vec N) = \int \mathrm{d}^3 x N^lH_l\end{equation}
we want to show that it generates infinitesimal spatial diffeomorphism (it's sufficient to do it on the phase space variables)
\begin{equation}
\left\lbrace q_{ab}(y),H(\vec N)\right\rbrace = \int \mathrm{d}^3 x N^l \frac{\delta H_l(x)}{\delta \pi^{ab} (y)} = 2 \nabla_{(a}N_{b)} = L_{\vec{N}} q_{ab}
\end{equation}
in fact from the definition of the Lie derivatives
\begin{equation*} 
\mathcal L_{\vec{N}} q_{ab} = (\nabla_aN^c)q_{cb} + (\nabla_bN^c)q_{ac} - \underbrace{N^c\nabla_c q_{ab}}_{=0} = 2 \nabla_{(a}N_{b)}
\end{equation*}
Note that the functional derivative is easy to compute with a by part integration. The last term vanishes for metric compatibility and in the Lie derivative you can substitute $\partial$ with $\nabla$.
\begin{align*}
\left\lbrace \pi^{ab},H(\vec N)\right\rbrace &= \int -N^l \frac{\delta H_l(x)}{\delta q_{ab} (y)} = -2 \pi^{ca}\nabla_cN^b + \sqrt{q} \nabla_k\left( \frac{N^k}{\sqrt{q}} \pi^{ab}\right)  = \\
&=-2 \pi^{ca}\nabla_cN^b \frac{\sqrt q }{\sqrt q} + \sqrt q (\nabla_kN^k) \frac{\pi^{ab}}{\sqrt q} + \sqrt q N^k  \nabla_k\frac{\pi^{ab}}{\sqrt{q}} = \\
&=\sqrt{q} \mathcal L_{\vec N} \frac{\pi^{ab}}{\sqrt q} + \frac{\pi^{ab}}{\sqrt q} (\nabla_kN^k) \sqrt{q}  = \nonumber \\
&=\sqrt{q} \mathcal L_{\vec N} \frac{\pi^{ab}}{\sqrt q} + \frac{\pi^{ab}}{\sqrt q} \mathcal L_{\vec N}\sqrt{q} \equiv \mathcal L_{\vec{N}} \pi^{ab}
\end{align*}
Note that $\pi^{ab}$  is a tensorial density so its lie derivative has to be defined as above. Given $A$ an arbitrary function of $q$ and $\pi$ using Leibniz we can compute the Poisson bracket easily
\begin{align*}
\left\lbrace A,H(\vec N)\right\rbrace &= \int \frac{\delta A}{\delta q_{ab}}\frac{\delta H(\vec N)}{\delta \pi^{ab}} - \frac{\delta H(\vec N)}{\delta q_{ab}}\frac{\delta A}{\delta \pi^{ab}} = \\
&=\int \frac{\delta A}{\delta q_{ab} }\left\lbrace q_{ab} ,H(\vec N)\right\rbrace - \left\lbrace \pi^{ab} ,H(\vec N)\right\rbrace \frac{\delta A}{\delta \pi^{ab}}=\\
&=\int \frac{\delta A}{\delta q_{ab}}\mathcal L_{\vec{N}} q_{ab} - \mathcal L_{\vec{N}} \pi^{ab}  \frac{\delta A}{\delta \pi^{ab}} =  \mathcal L_{\vec{N}} A
\end{align*}
In particular for $H$ and $H_a$ as function of $h$ and $\pi$ we deduce
\begin{align*}
\mathcal L_{\vec{N}} H_b &= \sqrt{q} \mathcal L_{\vec{N}} \frac{H_b}{\sqrt q} + \frac{H_b}{\sqrt q} \mathcal L_{\vec{N}}\sqrt{q} =\\
&=\sqrt q N^a \partial_a \frac{H_b}{\sqrt q} + H_a \partial_bN^a + H_b \nabla_aN^a=\\
&= N^a\partial_aH_b + N^aH_b \sqrt q \partial_a \frac{1}{\sqrt{q}} + H_a \partial_bN^a + H_b \nabla_aN^a=\\
&= N^a\partial_aH_b - N^aH_b \partial_a(\sqrt q) \frac{1}{\sqrt{q}} + H_a \partial_bN^a + H_b \nabla_aN^a=\\
&= N^a\partial_a H_b + H_a \partial_b N^a + H_b (\nabla_aN^a -\frac{ N^a}{\sqrt q}\partial_a \sqrt q)=\\
&= N^a\partial_a H_b + H_a \partial_b N^a + H_b\partial_a N^a
\end{align*}
In fact 
\begin{equation*}
\nabla_aN^a = \partial_a N^a + \Gamma ^a _{ba}N^b \end{equation*}
\begin{equation*}\frac{ N^a}{\sqrt q}\partial_a \sqrt q = \frac{1}{2}N^a q^{ab}\partial _a q_{ab} = \Gamma ^a _{ba}N^b\end{equation*}
Where the last equality is due to metric compatibility. We can conclude that
\begin{equation}
\left\lbrace \int N^aH_a, H_b\right\rbrace  = -N^a\partial_a H_b - H_a \partial_b N^a - H_b\partial_a N^a
\end{equation}
That is the smeared version of
\begin{equation}\left\lbrace H_a(x), H_b(y)\right\rbrace = H_a (y)\partial _b \delta(x-y) - H_b(x)\partial '_a\delta(x-y)\end{equation}
Moreover also the second Poisson bracket is easy to compute using the fact that $H(\vec N)$ is the generator of infinitesimal diffeomorphism along $\vec{N}$.
\begin{align}
\mathcal L_{\vec{N}} H &= \sqrt{q} \mathcal L_{\vec{N}} \frac{H}{\sqrt q} + \frac{H}{\sqrt q} \mathcal L_{\vec{N}}\sqrt{q} =\\
&=\sqrt q N^a \partial_a \frac{H}{\sqrt q} + \frac{H}{\sqrt{q}} (\frac{1}{2}\sqrt q q^{cd}\mathcal L_{\vec{N}}q_{cd})=\nonumber\\
&=N^a\partial_a H + \sqrt{q}N^aH\partial_a \frac{1}{\sqrt{q}} + \frac{H}{\sqrt{q}}\frac{2\sqrt{q}}{2}q^{cd}q_{ca}\nabla_d N^a=\nonumber\\
&=N^a\partial_a H + H \partial_a N^a + \sqrt{q} N^a H (\frac{-1}{2})\frac{1}{\sqrt{q}} q^{cd} \partial_a q_{cd} + H \Gamma^a_{ba}N^b  =\nonumber\\
&=N^a\partial_a H + H \partial_a N^a + \sqrt{q} N^a H (\frac{-1}{2})\frac{1}{\sqrt{q}} q^{cd} \underbrace{\nabla_a q_{cd}}_{=0} = N^a\partial_a H + H \partial_a N^a \nonumber
\end{align}
putting all together
\begin{equation}
\label{trucco2}
\left\lbrace \int N^aH_a, H\right\rbrace  = -N^a\partial_a H - H \partial_a N^a
\end{equation}
that is equivalent to
\begin{equation}
\left\lbrace H_a(x), H(y)\right\rbrace  = H\partial_a \delta(x-y)
\end{equation}

To ultimate the computation of the second Poisson bracket we need some preliminary results. Given an arbitrary function  $f(\pi,h)$ 
\begin{equation}
\left\lbrace \left\lbrace f,H(\vec N_1)\right\rbrace,H(\vec N_2) \right\rbrace  = \mathcal L_{\vec {N_1}} \mathcal L_{\vec{N_2}} f
\end{equation}
using the Jacobi identity
\begin{align}\left\lbrace f,\left\lbrace H(\vec N_1),H(\vec N_2)\right\rbrace \right\rbrace &
=\left\lbrace \left\lbrace H(\vec N_2),f\right\rbrace ,H(\vec N_1)\right\rbrace  + \left\lbrace \left\lbrace f,H(\vec N_1)\right\rbrace ,H(\vec N_2)\right\rbrace=\nonumber\\
&= \left(-\mathcal L_{\vec{N_1}}\mathcal L_{\vec{N_2}} + \mathcal L_{\vec{N_2}}\mathcal L_{\vec{N_1}} \right)f = \left[ \mathcal L_{\vec{N_2}},\mathcal L_{\vec{N_1}}\right] f = \nonumber\\
&=\mathcal L_{\left[\vec{N_2},\vec{N_1}\right]}f  = \left\lbrace f,\int \left[N_2,N_1\right]^aH_a\right\rbrace \nonumber
\end{align}
If we smear the \eqref{trucco2} we obtain:
\begin{equation}
\left\lbrace \int N_1^aH_a,\int N H \right\rbrace = \int \mathcal L_{\vec N}N H
\end{equation}
\begin{equation}
\left\lbrace\int N_1^aH_a,\int N_2^aH_a \right\rbrace  = \int \left[N_1,N_2\right]^aH_a
\end{equation}
Finally we compute the last Poisson bracket, if $N\neq N'$ (if they are equal it vanish)
\begin{equation}
\left\lbrace H(N),H(N')\right\rbrace = \int \mathrm{d}^3 x \frac{\delta H(N)}{\delta q_{ab} (x)}\frac{\delta H(N')}{\delta \pi^{ab} (x)} - \frac{\delta H(N')}{\delta q_{ab} (x)}\frac{\delta H(N)}{\delta \pi^{ab} (x)}\end{equation}
We notice that all the algebraical terms in $q$ and $\pi$ simplify each other. In fact 
\begin{equation}
\left. \frac{\delta H(x)}{\delta q_{ab} (y)}\right|_{\substack{algebraic\\ part}} = f^h(x)_{ab}\delta(x-y)
\end{equation}
\begin{equation}
\left. \frac{\delta H(x)}{\delta \pi_{ab} (y)}\right|_{\substack{algebraic\\ part}} = f_\pi(x)^{ab}\delta(x-y)
\end{equation}
\begin{equation}
\left\lbrace \left.H(N)\right|_{\substack{algebraic\\ part}},\left.H(N')\right|_{\substack{algebraic\\ part}}\right\rbrace = \int \mathrm{d}^3 x Nf^h_{ab}N'f_\pi^{ab} -N'f_\pi^{ab}Nf^h_{ab} = 0\end{equation}
The non algebraic part in $q$ are in $R$ conversely the Hamiltonian constraint is algebraic in $\pi$.
\begin{equation}
\label{mostro}
\left\lbrace H(N),H(N')\right\rbrace = \int \mathrm{d}^3 x \frac{\delta H(N){\substack{non\\ algebraic\\ term}}}{\delta q_{ab} (x)}\frac{\delta H(N')}{\delta \pi^{ab} (x)} - \frac{\delta H(N'){\substack{non \\algebraic\\ term}}}{\delta q_{ab} (x)}\frac{\delta H(N)}{\delta \pi^{ab} (x)}\end{equation}
We compute (ignoring algebraic terms for simplicity)
\begin{align*}
\delta_q \int N H &= - \int N \sqrt{q} q^{ab}\delta R_{ab}=\\
&=-\int N \sqrt{q} \left( \nabla_a\nabla_b\delta q_{cd}q^{ac}q^{bd} - q^{cd}\nabla^b\nabla_b \delta q_{cd} \right) =\\
&=-\int \delta q_{cd} \sqrt{q}\left( \nabla^c\nabla^d N - \nabla^2 (N q^{cd})\right)
\end{align*}
Substituting this variation in the Poisson bracket the first term becomes
\begin{align}
\int \frac{\delta \int NH}{\delta q_{ab}}\frac{\delta \int N' H}{\delta \pi^{ab}} &= -\int \sqrt{q} \left(  \nabla^a\nabla^b N - \nabla^2 (N q^{ab})\right)2N'\left(\frac{\pi_{ab}}{\sqrt{q}} - \frac{\pi q_{ab}}{\sqrt{q}(3-1)} \right)=\nonumber\\
&= - 2\int (\nabla^a\nabla^b N)\pi_{ab}N' - (\nabla^2N) \pi N' - (\nabla^2N) N' \frac{\pi}{3-1} + (\pi \nabla^2N) \frac{q^{ab}q_{ab}}{3-1}N' = \nonumber\\
&= 2\int (\nabla^a\nabla^b N)\pi_{ab}N'  + \left( \frac{3}{3-1} - \frac{1}{3-1} -1\right) \pi N' \nabla^2 N =\nonumber\\
&=- 2\int \sqrt{q} (\nabla^a\nabla^bN)\frac{\pi_{ab}}{\sqrt{q}}N' = 2 \int \sqrt{q} \nabla^bN \nabla^a \left( \frac{\pi_{ab} N'}{\sqrt{q}}\right) 
\end{align}
summing also the second term
\begin{eqnarray}
&&[\ref{mostro}] =  2 \int \sqrt{q}\left[  \nabla^bN \nabla^a \left( \frac{\pi_{ab} N'}{\sqrt{q}}\right) - \nabla^bN' \nabla^a \left( \frac{\pi_{ab} N}{\sqrt{q}}\right)\right] = \\
&& = 2\int \sqrt{q} \left[ \nabla^bN \nabla^aN' \frac{\pi_{ab}}{\sqrt{q}} + N'(\nabla^bN) \nabla^a\left( \frac{\pi_{ab}}{\sqrt{q}}\right)   - \nabla^bN' \nabla^aN \frac{\pi_{ab}}{\sqrt{q}} + N(\nabla^bN') \nabla^a\left( \frac{\pi_{ab}}{\sqrt{q}}\right)  \right] =\nonumber\\
&& =\int \left( -N'\nabla^bN + N\nabla^b N'\right) 2 \sqrt{q} \nabla^a\left( \frac{\pi_{ab}}{\sqrt{q}}\right) = \int \left( N\nabla^b N'-N'\nabla^bN \right) H_a\nonumber
\end{eqnarray}
\end{computation}

\section{Tetrad formulation}

A tetrad is a quadruple of 1-forms, $e^I_\mu(x)$, $I=0,1,2,3$ such that
\begin{equation}
g_{\mu\nu}(x)= e^I_\mu(x) e^J_\nu(x) \eta_{IJ}.
\end{equation}
By its definition, it provides a local isomorphism between a general reference frame and an inertial one, characterized by the flat metric $\eta_{IJ}$. A local inertial frame is defined up to a Lorentz transformation, and in fact notice that the definition is invariant under
\begin{equation}
e^I_\mu(x) \longrightarrow \tilde e_\mu^I(x) = \Lambda^I_J(x) e^J_\mu(x).
\end{equation}
This means that the ``internal'' index $I$ carries a representation of the Lorentz group. Contracting vectors and tensors in spacetime with the tetrad, we get objects that transform under the Lorentz group, e.g. $e^I_\mu n^\mu = n^I$. The tetrad thus provides an isomorphism between the tangent bundle of $\cal M$,  $T(\mathcal M) = \bigcup_\rho \left(\rho,T_\rho(\mathcal M)\right)$, and a Lorentz principal bundle $F=\left( \mathcal M, SO(3,1)\right)$.
On this bundle we have a connection $\omega^{IJ}_\mu$, that is a 1-form with values in the Lorentz algebra, which we can use to define covariant differentiation of the fibres, 
\begin{equation}
D_\mu v^I(x) = \partial_\mu v^I (x) + \omega_\mu^{I}{}_{J}(x) v^J(x).
\end{equation}
This is the analogue of the covariant derivative $\n_\mu = \p_\mu + \Gamma_\mu$ for vectors in $T(\mathcal M)$.
We can also define the derivative for objects with both indices, such as the tetrad,
\be
{\cal D}_\mu e^I_\nu = \p_\mu e^I_\nu + \om^I_\mu{}_J e^J_\nu - \Gamma^\rho_{\nu\mu} e^I_\rho.
\ee
As the Levi-Civita connection $\Gamma(g)$ is metric-compatible, i.e. $\n_\mu g_{\nu\rho}=0$, we require $\om_\mu$ to be tetrad-compatible, i.e. ${\cal D}_\mu e^I_\nu\equiv 0$, and call it spin connection.
This implies
\be
\p_{(\mu} e^I_{\nu)} + \om^I_{(\mu}{}_J e^J_{\nu)} = \Gamma^\rho_{(\nu\mu)} e^I_\rho, \qquad
\p_{[\mu} e^I_{\nu]} + \om^I_{[\mu}{}_J e^J_{\nu]} = \Gamma^\rho_{[\nu\mu]} e^I_\rho \equiv 0,
\ee
where we separed the spacetime indices into their symmetric and antisymmetric combinations, and used the fact that the Levi-Civita connection $\Gamma(g)$ has no antisymmetric part.

From these equations we immediately obtain the following relation between the spin and Levi-Civita connections,
\begin{equation}
\label{oe}
\omega^I_{\mu J} = e^I_\nu \nabla_\mu e^\nu_J,
\end{equation}
as well as the fact that the spin connection satisfies
\begin{equation}
\mathrm d_\omega e^I = \mathrm d e^I + \omega^I{}_J \wedge e^J = \left(\partial_\mu e_\nu^I + \omega^I_{\mu J} e^J_\nu \right)\mathrm d x^\mu \wedge \mathrm d x^\nu = 0.
\end{equation}
This equation is known as Cartan's first structure equation. Here we introduced the exterior calculus of forms, with d the exterior derivative, $\mathrm d_\om$ the covariant exterior derivative, and $\w$ the wedge product. See \cite{IshamBook} for an introduction to this formalism.

Given the connection, we define its curvature
\begin{equation}
F^{IJ} = \mathrm d \omega^{IJ} + \omega^I{}_K \wedge \omega^{KJ},
\end{equation}
whose components are
\begin{equation}\label{Fcomps}
F^{IJ}_{\mu \nu} = \partial_\mu \omega_\nu^{IJ} - \partial_\nu \omega_\mu^{IJ} + \omega^{I}{}_{K\mu} \omega^{KJ}_\nu -\omega^{J}{}_{K\mu} \omega^{KI}_\nu.
\end{equation}
Using the solution $\omega(e)$ given in \Ref{oe}, an explicit calculation gives
\begin{equation}
\label{cartan2}
F_{\mu \nu}^{IJ}\left(\omega\left(e\right)\right) \equiv e^{I\rho} e^{J \sigma}R_{\mu \nu \rho\sigma}(e),
\end{equation}
where $R_{\mu\nu\rho \sigma}(e)$ is the Riemann tensor constructed out of (the metric defined by) the tetrad $e^I_\mu$. This relation is known as Cartan second structure equation. It shows that general relativity is a gauge theory whose local gauge group is the Lorentz group, and the Riemann tensor is nothing but the field-strength of the spin connection.

\begin{computation}{Cartan second structure equation}
Starting from the definition of $F$ in \Ref{Fcomps} and inserting \Ref{oe}, we have
\begin{align*}
F^{IJ}_{\mu\nu} =& \p_\mu e^I_\rho \partial_\nu e^{\rho J} + \partial_\mu e^I_\rho \Gamma^{\rho}_{\sigma \nu} e^{\sigma J} + e^I_\rho \partial_\mu\left(  \Gamma^{\rho}_{\sigma \nu} \right) e^{\sigma J} + e^{I}_\rho  \Gamma^{\rho}_{\sigma \nu} \partial_\mu e^{\sigma J}+ e^I_\rho \partial_\mu e^\rho_K e^K_\sigma \partial_\nu e^{\sigma J}+\\
&+ e^I_\rho \Gamma^{\rho}_{\delta \mu}e^\delta_K e^K_\sigma \partial_\nu e^{\sigma J} + e^I_\rho \partial_\mu e^\rho_K e^K_\sigma \Gamma^{\sigma}_{\delta \nu} e^{\delta J} + e^I_\rho \Gamma^{\rho}_{\delta \mu}e^\delta_K e^K_\sigma  \Gamma^{\sigma}_{\eta \nu} e^{\eta J} - (\mu \leftrightarrow \nu).
\end{align*}
Next, we use $e^I_\rho e^\rho_K=\delta^I_K$ and $e^I_\rho \partial_\mu e^\rho_K = -\partial_\mu (e^I_\rho)  e^\rho_K$ to rewrite this expression as 
\begin{align*}
F^{IJ}_{\mu\nu} &= e^I_\rho  e^{\sigma J} \partial_\mu\left(  \Gamma^{\rho}_{\sigma \nu} \right) + e^I_\rho  e^{\sigma J} \Gamma^{\rho}_{\delta \mu} \Gamma^{\delta}_{\sigma \nu} + \partial_mu e^I_\rho \partial_\nu e^{\rho J} - \partial_\mu e^I_\rho \partial_\nu e^{\rho J}+\\
& + \partial_\mu e^I_\rho \Gamma^{\rho}_{\sigma \nu} e^{\sigma J} -  \partial_\mu e^I_\rho  \Gamma^{\rho}_{\sigma \nu} e^{\sigma J} +  e^{I}_\rho  \Gamma^{\rho}_{\sigma \nu} \partial_\mu e^{\sigma J}+ e^I_\rho \Gamma^{\rho}_{\sigma \mu} \partial_\nu e^{\sigma J} - 
(\mu \leftrightarrow \nu) \\
& = 2 e^{I}_\rho  e^{J\sigma} \left( \partial_{(\mu}  \Gamma^{\rho}_{\sigma \nu)}  + \Gamma^{\rho}_{\delta (\mu} \Gamma^{\delta}_{\sigma \nu)} \right) =  e^{I\rho}  e^{\sigma J} R_{\mu \nu \rho\sigma}.
\end{align*}
\lightline\\
\noindent \emph{Relation between the determinants}.
The determinant $g$ of $g_{\mu\nu}$ is related to the determinant of the tetrad $e$ by the simple relation
\begin{equation}
\label{det}
g = - e^2.
\end{equation}
This expression can be easily derived recalling Caley's formula for the determinant of a matrix, 
\begin{equation}
g = \mathrm{det} g_{\mu \nu} = \frac{1}{4!} \varepsilon^{\mu\nu\rho\sigma}\varepsilon^{\alpha\beta\gamma\delta} g_{\mu\alpha}g_{\nu\beta} g_{\rho\gamma}g_{\sigma\delta}.
\end{equation}
If we substitute the expression of $g_{\mu\nu}$ in terms of tetrads we get
\begin{align*}
g &= \mathrm{det}\left( e_\mu^I e_\nu^J \eta_{IJ} \right) = \frac{1}{4!} \varepsilon^{\mu\nu\rho\sigma}\varepsilon^{\alpha\beta\gamma\delta}  e_\mu^I e_\alpha^J \eta_{IJ}  e_\nu^K e_\beta^L \eta_{KL}  e_\rho^M e_\gamma^N \eta_{MN}   e_\sigma^O e_\delta^P \eta_{OP} = \\
&=\frac{1}{4!} \varepsilon^{\mu\nu\rho\sigma} e_\mu^Ie_\nu^K  e_\rho^Me_\sigma^O  \varepsilon^{\alpha\beta\gamma\delta} e_\alpha^J e_\beta^L  e_\gamma^Ne_\delta^P \eta_{IJ}  \eta_{KL} \eta_{MN}  \eta_{OP} = \\
&=\frac{1}{4!} e^2 \varepsilon^{IKMO} \varepsilon^{JLNP} \eta_{IJ}  \eta_{KL} \eta_{MN}  \eta_{OP} = - e^2
\end{align*}

\end{computation}

\subsection{The action in terms of tetrads}
The Einstein-Hilbert action can be rewritten as a functional of the tetrad in the following way (recall we take units $16\pi G=1$),
\begin{equation}\label{EHtet}
S_{EH}(e^I_\mu) =  \f12 \varepsilon_{IJKL} \int e^I \wedge e^J \wedge F^{KL}\left(\omega(e)\right).
\end{equation}
On top of the invariance under diffeomorphism, this reformulation of the theory possesses an additional gauge symmetry under local Lorentz transformations.

\begin{computation}{Rewriting the actions in terms of tetrads}
Explicitly using (\ref{cartan2}) and (\ref{det})
\begin{align*}
S_{EH}(g_{\mu\nu}(e)) &= \int \mathrm d^4x\, \sqrt{-g} \,g^{\mu \nu} R_{\mu \nu} = \int \mathrm d^4x\, e\, e^\mu_I e^{\nu I} R_{\mu \rho \nu \sigma} e^\rho_J e^{\sigma J} =\\
&= \int \mathrm d^4x e e^\mu_I e_J^\rho F^{IJ}_{\mu \rho}\left(\omega(e)\right) = \int \mathrm d^4x \frac{1}{4} \varepsilon_{IJKL}\varepsilon^{\mu\rho\alpha\beta}e^K_\alpha e^L_\beta F^{IJ}_{\mu \rho}\left(\omega(e)\right)  \\
&= \int \frac{1}{2} \varepsilon_{IJKL} e^I \wedge e^J \wedge F^{KL}\left(\omega(e)\right)
\end{align*}

\end{computation}
A fact which plays an important role in the following is that we can lift the connection to be an \emph{independent variable}, and consider the new action
\begin{equation}
S(e^I_\mu,\om^{IJ}_\mu) = \f12 \varepsilon_{IJKL} \int e^I \wedge e^J \wedge F^{KL}\left(\omega\right).
\end{equation}
Although it depends on extra fields, this action remarkably gives the same equations of motion as the Einstein-Hilbert one \Ref{EHtet}.
This happens because the extra field equations coming from varying the action with respect to $\om$ do not add anything new: they simply impose the form \Ref{oe} of the spin connection, and general relativity is thus recovered.

\begin{computation}{Deriving the fields equations}
One first verifies the Palatini identity
\begin{equation}\label{EC}
 \delta_\omega F^{KL}\left(\omega\right) = \mathrm d_\omega \delta \omega^{KL}.
\end{equation}
Then, the variation gives 
\begin{equation}
\delta_\omega S = \frac{1}{2} \varepsilon_{IJKL} \int e^I \wedge e^J \wedge \mathrm d_\omega \delta \omega^{KL}
= - \frac{1}{2} \varepsilon_{IJKL}  \int \mathrm d_\omega \left( e^I \wedge e^J\right) \wedge  \delta \omega^{KL}
\end{equation}
after an integration by part. Imposing the vanishing of the variation, we obtain the field equation
\begin{equation}
\label{fieldeq1}
\varepsilon_{IJKL}e^I \wedge \mathrm d_\omega e^J = 0. 
\end{equation}
If the tetrad is invertible this equation implies $d_\omega e^J = 0$, which in turns implies \Ref{oe} in terms of the Levi-Civita connection of the metric associated with $e^I_\mu$.

\end{computation}

As it gives the same field equations, \Ref{EC} can be used as the action of general relativity. Notice that only first derivatives appears, thus it provides a first order formulation of general relativity. Furthermore, the action is polynomial in the fields, a desiderable property for quantization. On the other hand, there are two non-trivial aspects to take into account:
\begin{itemize}
\item The equivalence with general relativity holds only if the tetrad is \emph{non-degenerate}, i.e. invertible. On the other hand, \Ref{EC} is also defined for degenerate tetrads, since inverse tetrads never appear. Compare the situation with the Einstein-Hilbert action, where the inverse metric appears explicitly. Hence, the use of \Ref{EC} leads naturally to an extension of general relativity where a sector with degenerate tetrads, and thus degenerate metrics, exists.
\item If we insist on the connection being an independent variable, there exists a second term that we can add to the Lagrangian that is compatible with all the symmetries and has mass dimension 4:
\begin{equation}
\delta_{IJKL} e^I \wedge e^J \wedge F^{KL}(\om),
\end{equation}
where $\delta_{IJKL} \equiv  \delta_{I[K} \delta_{L]J}$. 
This term is not present in the ordinary second order metric, since when \Ref{oe} holds,
\begin{equation}
\delta_{IJKL} e^I \wedge e^J \wedge F^{KL}(\om(e)) = \eps^{\mu\nu\rho\sigma} R_{\mu\nu\rho\sigma}(e)\equiv 0.
\end{equation}
\end{itemize}
Adding this second term to \Ref{EC} with a coupling constant $1/\gamma$ leads to the so-called Holst action \cite{Holst:1995pc}
\begin{equation}
\label{Holst}
S\left(e,\omega\right) = \left(\frac{1}{2} \varepsilon_{IJKL}+\frac{1}{\gamma} \delta_{IJKL}\right) \int e^I \wedge e^J \wedge F^{KL}\left(\omega\right).
\end{equation}
Assuming non-degenerate tetrads, this action leads to the same field equations of general relativity,
\begin{equation}
\omega_{\mu}^{IJ} = e^I_\nu \nabla_\mu e^{J\nu}, \qquad G_{\mu \nu}(e)=0.
\end{equation}
This result is \emph{completely independent} of the value of $\gamma$, which is thus a parameter irrelevant in classical vacuum general relativity. It will however turn out to play a key role in the quantum theory, where it is known as the Immirzi parameter.\footnote{The Immirzi parameter becomes relevant also at the classical level if source of torsion are present \cite{RovelliTors,FreidelTors}.}

\begin{computation}{Deriving the fields equations for the Holst action}
The vanishing of the variations give
\bea
&& \left(\f12 \varepsilon_{IJKL}+\frac{1}{\gamma} \delta_{IJKL}\right) e^I \wedge \mathrm d_\omega e^J = 0, 
\\
&& \left(\f12 \varepsilon_{IJKL}+\frac{1}{\gamma} \delta_{IJKL}\right) e^J \wedge  F^{KL}\left(\omega\right) = 0.
\eea
For invertible tetrads, the first one is again uniquely solved by \Ref{oe}. Substituting this solution into the second equation we get
\begin{align*}
& e^{I\nu}G^\alpha_{\nu} +\frac{1}{\gamma} \delta_{IJKL} \varepsilon^{\mu \sigma \delta \alpha} e^J_\mu F^{KL}_{\sigma \delta}= e^{I\nu}G^\alpha_{\nu}+\frac{1}{\gamma} \delta_{IJKL} \varepsilon^{\mu \sigma \delta \alpha} e^J_\mu  e^{K\nu} e^{L\rho} R_{\sigma \delta \nu \rho} = \\
&=e^{I\nu}G^\alpha_{\nu} + \frac{1}{2 \gamma} \varepsilon^{\mu \sigma \delta \alpha}\left(e^{I\nu}\delta^\rho_\mu - e^{I\rho}\delta^\nu_\mu \right)R_{\sigma \delta \nu \rho} = 
e^{I\nu}G^\alpha_{\nu} - \frac{1}{\gamma} \varepsilon^{\sigma \delta \mu \alpha}e^{I\nu}R_{\sigma \delta \mu \nu} = 
e^{I\nu}G^\alpha_{\nu} = 0,
\end{align*}
which is equivalent to $G_{\mu\nu}=0$ thanks to the non-degeneracy of the tetrad. In the last step we used the first Bianchi identity $\varepsilon^{\sigma \delta \mu \alpha}R_{\sigma \delta \mu \nu} = 0$.

\end{computation}

\subsection{Hamiltonian analysis of tetrad formulation}

For the Hamiltonian formulation we proceed as before, assuming a $3+1$ splitting of the space-time ($\mathcal M \cong \mathbb{R}\times \Sigma$) and coordinates $(t,x)$. We introduce the lapse function and the shift vector ($N$, $N^a$) and the ADM decomposition of the metric \Ref{gADM}. It is easy to see that a tetrad for the ADM metric is given by 
\be
e_0^I = e^I_\mu \tau^\mu = N n^I + N^a e_a^I, \qquad \d_{ij} e_a^i e^j_{b} = g_{ab}, \ i=1,2,3.
\ee
The ``triad'' $e^i_a$ is the spatial part of the tetrad. 
As before, we want to identify canonically conjugated variables and perform the Legendre transform, but we now have two new features which complicate the analysis. The first one is the tetrad formulation, which in particular has introduced a new symmetry in the action: the invariance under local Lorentz transformations. 
As a consequence, we expect more constraints to appear, corresponding to the generators of the new local symmetry. The second one is the use of the tetrad and the connection as independent fields. Therefore, the conjugate variables are now functions of both $e^I_a$ and $\om^{IJ}_a$ (and their time derivatives), as opposed to be functions of the metric $g_{ab}$ only. 

The consequence of these novelties is a much more complicated structure than in the metric case. In particular, the constraint algebra is \emph{second class}.
However, there is a particular choice of variables which simplifies the analysis, making it possible to implement a part of the constraint and reducing the remaining ones to first class again. These are the famous Ashtekar variables, which we now introduce.\footnote{On the general analysis with the second class constraints see \cite{BarroseSa:2000vx}.}

To simplify the discussion, it is customary to work in the ``time gauge'' $e^I_\mu n^\mu=\d^I_0$, where 
\begin{equation}
e^0_\mu = (N , 0 ) \longrightarrow e^I_0 = \left( N, N^ae_a^i \right).
\end{equation}
The crucial change of variables is the following: we define the \emph{densitized triad}
\begin{equation}\label{defE}
E_i^a = e e_i^a = \frac{1}{2} \varepsilon_{ijk}\varepsilon^{abc} e^j_b e^k_c,
\end{equation}
and the \emph{Ashtekar-Barbero connection}
\begin{equation}\label{defA}
A^i_a = \gamma \omega_a^{0i} + \frac{1}{2} \varepsilon^i_{jk} \omega_a^{jk}.
\end{equation}
These variables turns out to be conjugated. In fact, we can rewrite the action (\ref{Holst}) in terms of the new variables as \cite{BarroseSa:2000vx, ThiemannBook}
\begin{equation}\label{AHam}
S(A,E,N,N^a) = \frac{1}{\gamma} \int \mathrm d t \int_\Sigma \mathrm d^3 x \left[\dot A^i_a E^a_i - A^i_0 D_aE^a_i - N H - N^aH_a\right],
\end{equation}
where
\begin{align}
&G_j\equiv D_aE^a_i =  \partial_aE^a_j + \varepsilon_{jk\ell}A^j_a E^{a\ell}, \label{Gauss}\\ 
&H_a=\frac{1}{\gamma}F^j_{ab}E^b_j - \frac{1 + \gamma^2}{\gamma} K^i_a G_i, \\
&H =\left[F^j_{ab} - \left( \gamma^2+1\right) \varepsilon_{jmn}K^m_aK^n_b\right] \frac{\varepsilon_{jk\ell}E^a_kE^b_\ell}{\mathrm{det} E} + \dfrac{1 + \gamma^2}{\gamma} G^i \partial_a \frac{E^a_i}{\mathrm{det} E} \label{HamiltonianConstraint}.
\end{align}
The resulting action is similar to \Ref{EHaction2}, with $(A,E)$ as canonically conjugated variables, as opposed to $(q,\pi)$. Lapse and shift are still Lagrange multipliers, and consistently we still refer to $H(A,E)$ and $H_a(A,E)$ as the Hamiltonian and space-diffeomorphism constraints. 

The algebra is still first class. The new formulation in terms of tetrads has introduced the extra constraint \Ref{Gauss}. The reader familiar with gauge theories will recognize it as the Gauss constraint. Just as the $H^\mu$ constraints generate diffeomorphisms, the Gauss constraint generates gauge transformations. 
It is in fact easy to check that $E^b_j$ and $A^i_a$ transform respectively as an SU(2) vector and as an SU(2) connection under this transformation.

\begin{computation}{The algebra generated by the Gauss constraint}
We define the smearing of the Gauss constraint
\begin{equation}
G\left(\Lambda\right) = \int \mathrm d^3 x G_i (x) \Lambda^i(x).
\end{equation}
Its Poisson bracket with the canonical variables give
\begin{align*}
\left\lbrace\int \mathrm d^3 x  \Lambda^j(x) G_j (x), E^a_i(y) \right\rbrace &= \int\mathrm d^3 x  \Lambda^j(x) \left\lbrace\partial_bE^b_j + \varepsilon_{mjn}A^m_b E^{bn}, E^a_i(y) \right\rbrace=\\
& =\gamma \varepsilon_{mjn} \Lambda^j(y) E^{bn}(y) \delta^a_b \delta^m_i =  \gamma \varepsilon_{ijn} \Lambda^j(y) E^{an}(y), \\
\end{align*}
\begin{align*}
 \left\lbrace\int \mathrm d^3 x  \Lambda^j(x) G_j (x), A_a^i(y) \right\rbrace &= \int\mathrm d^3 x  \Lambda^j(x) \left\lbrace\partial_bE^b_j + \varepsilon_{mjn}A^m_b E^{bn}, A_a^i(y) \right\rbrace=\\
& =\gamma \partial_a\Lambda^i(y) + \gamma \varepsilon_{mji}\Lambda^j(y)A^m_a(y). 
\end{align*}
Hence,
\begin{align*}
&\left\lbrace G(\Lambda), \partial_aE^a_i(y) + \varepsilon_{jik}A^j_a E^{ak}(y) \right\rbrace= \\ &=  \gamma \varepsilon_{ijn} \partial_a \left(\Lambda^j(y) E^{an}(y)\right) - \gamma \varepsilon_{jik}A^j_a(y)  \varepsilon_{k\ell n} \Lambda^\ell(y) E^{an}(y) +\\
&+\gamma \varepsilon_{jik} E^{ak}\left(\partial_a\Lambda^j(y) + \varepsilon_{m\ell j}\Lambda^\ell(y)A^m_a(y) \right)=\\
&=\gamma \varepsilon_{ijn} \Lambda^j(y) \partial_a E^{an}(y) + \gamma \left(\varepsilon_{ijk}  \varepsilon_{\ell nk} - \varepsilon_{ink} \varepsilon_{\ell j k}\right) \Lambda^\ell(y) E^{an}(y) A^j_a(y)=\\
&=\gamma \varepsilon_{i\ell k} \Lambda^\ell(y) \partial_a E^{ak}(y) + \gamma  \varepsilon_{i\ell k}  \Lambda^\ell(y)  \varepsilon_{j kn} E^{an}(y) A^j_a(y)=\gamma \varepsilon_{i\ell k}\Lambda^\ell(y)  G_k (y).
\end{align*}
Finally, smearing also the second term in the above bracket we get
\begin{equation}
\left\lbrace G(\Lambda_1), G(\Lambda_2)\right\rbrace = \frac{\gamma}{2} G\left(\left[\Lambda_1,\Lambda_2\right]\right),
\end{equation}
which we recognize as the $su(2)$ algebra structure equations.\\
\end{computation}

We should not be surprised of the appearance of this extra constraint. When we use the tetrad formalism, we introduce a new symmetry in the theory, the invariance under local gauge transformation. The Gauss constraint is there to enforce this invariance at the canonical level.
On the other hand, it might be more puzzling that although the local gauge invariance of the covariant action was the full Lorentz group, the Legendre transform \Ref{AHam} is only invariant under SU(2).

The origin of this puzzle lies precisely in the change of variables (\ref{defE}-\ref{defA}). The Ashtekar-Barbero connection is an $SU(2)$ connection, not a Lorentz connection. In particular, it is not the pull-back of the space-time Lorentz connection \cite{Samuel}.
It should then be seen as an auxiliary variable, useful to recast the algebra in a first class form.

Summarizing, in this formulation of General Relativity the theory is described by an extended phase space  of dimension $18\cdot \infty^3$ with the fundamental Poisson bracket
\begin{equation}\label{AEalgebra}
\left\lbrace A^i_a(x), E^b_j(y) \right\rbrace = \gamma \delta^b_a \delta^i_j \delta^3(x,y) 
\end{equation}
This new internal index $i$ corresponds to the adjoint representation of SU$(2)$, and we can recover the old $12\cdot \infty^3$ dimensional phase space on the constraint surface $G_i=0$ dividing by gauge orbits generated by $G$.
This group SU$(2)$ should be seen as an auxiliary local symmetry group, since the connection with the original Lorentz group of the tetrad formulation is hidden in the change of variables (\ref{defE}-\ref{defA}). It is only for the special case $\gamma = i$ (which is the original one introduced by Ashtekar) that the relation is manifest: in that case, the SU$(2)$ corresponds to the self-dual subgroup of the Lorentz group. Notice that this case also leads to a simplification of the Hamiltonian constraint \eqref{HamiltonianConstraint}. On the other hand, the variables are now complex, and to recover general relativity one needs to impose reality conditions. These are particularly difficult to deal with at the quantum level, and for this reason most developments in LQG have focused on $\gamma$ real. This is what we do also in this review.

\subsection{Smearing of the algebra}
The next step is to smear the algebra \Ref{AEalgebra}, as we did previously with the ADM variables. This is needed in order to proceed with the quantization. At this stage, the different tensorial nature of $A_a$ and $E^a$ plays a key role. If we were doing gauge theory on flat spacetime, we would not pay attention to the different indices, and just smear the connection and the electric field with the same type of test functions. On the contrary, now that the non-trivial geometry of spacetime is our main goal, we have to be careful. In fact, a brief look at \Ref{defE} shows that the densitised triad is a 2-form. Hence, it is natural to smear it on a surface,
\begin{equation}
E_i(S)\equiv \int_S n_a E^a_i \mathrm{d}^2 \sigma, 
\ee
where $n_a = \varepsilon_{abc}\frac{\partial x^b}{\partial \sigma_1}\frac{\partial x^c}{\partial \sigma_2}$ is the normal to the surface. The quantity $E_i(S)$ is the \emph{flux} of $E$ across $S$. 

The connection on the other hand is a 1-form, so it is natural to smear it along a 1-dimensional path.
Recall that a connection defines a notion of parallel transport of the fiber over the base manifold. Consider a path $\gamma$ and a parametrization of it $x^a(s)\,: [0,1] \to \Sigma$. Given a connection $A^i_a$ we can associate to it an element of $SU(2)$ $A_a \equiv A_a^i \tau_i$ where $\tau_i$ are the generator of $SU(2)$.\\
\begin{minipage}[b]{0.2\linewidth}
\includegraphics[scale=1]{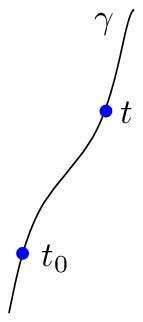} 
\end{minipage}
\begin{minipage}[b]{0.8\linewidth}
Then we can integrate $A_a$ along $\gamma$ as a line integral,
\begin{equation}
A^i_a \longrightarrow \int_\gamma A \equiv \int_0^1 \mathrm{d}s A^i_a(x(s)) \dfrac{\mathrm dx^a(s)}{\mathrm d s} \tau_i.
\end{equation}
Next, we define the holonomy of $A$ along $\gamma$ to be
\begin{equation}
h_\gamma = \mathcal P \exp\left( \int_\gamma A \right),
\end{equation}
\end{minipage}
where $\cal P$ stands for the path-ordered product. That is, 
\begin{equation}\label{holseries}
h_\gamma = \sum_{n=0}^\infty \ \ \iiint\limits_{1>s_n>\cdots >s_1>0}A(\gamma(s_1))\cdots A(\gamma(s_n)) ds_1 \cdots ds_n
\end{equation}
where we parametrized the line with $s\in[0,1]$.

\begin{computation}{On the definition of the holonomy}
More precisely, we call holonomy the solution of the differential equation 
\begin{equation}
\label{holonomyode}
\dfrac{d}{dt}h_\gamma(t) - h_\gamma(t)A(\gamma(t))=0, \qquad h_\gamma(0)=1.
\end{equation}
If we integrate the equation by iteration we have
\begin{align}
h_\gamma (t) &= 1 + \int_0^t A(\gamma(s))h_\gamma(s)ds =\\
& = 1 + \int_0^t A(\gamma(s_1))h_\gamma(s_1)ds_1 + \int_0^t\int_{s_1}^1 A(\gamma(s_1)) A(\gamma(s_2))h_\gamma(s_2)ds_1ds_2 \nonumber =
\\ & = \ldots \nonumber
\end{align}
therefore formally
\begin{equation}
h_\gamma(t) = \sum_{n=0}^\infty  \iiint\limits_{t>s_1>\cdots >s_n>0}A(\gamma(s_1))\cdots A(\gamma(s_n)) ds_1 \cdots ds_n.
\end{equation}
To complete the proof, one needs to show that the series is well defined. Indeed, it converges with respect to sup norm
\begin{equation}
\left|\left|h_\gamma\right|\right| \leq \sum_{n=0}^\infty \;\iiint\limits_{t>s_1>\cdots >s_n>0}\left|\left|A(\gamma(s_1))\cdots A(\gamma(s_n))\right|\right| ds_1 \cdots ds_n \leq \sum_{n=0}^\infty \left|\left|A\right|\right|^n \frac{t^n}{n!}.
\end{equation}
For further reference, let us also notice that the terms of the series can be written as integrals over square domains $(s_1,\ldots,s_n)\in[0,t]^n$, instead triangle domains $t>s_1>\cdots >s_n>0$. This gives
\begin{equation}\label{intmult}
h_\gamma(t)=\mathcal P \exp \left[ \oint_\gamma A\right] = \sum_{n=0}^\infty \frac{1}{n!} \iiint_{\square} \mathcal P \left( A(\gamma(s_1))\cdots A(\gamma(s_n)\right).
\end{equation}

\end{computation}

Let us list some useful properties of the holonomy.
\begin{itemize}
\item The holonomy of the composition of two paths is the product of the holonomies of each path,
\begin{equation}\label{prima}
h_{\beta\alpha} = h_{\beta} h_{\alpha}.
\end{equation}
\item Under a local gauge transformations $g(x)\in$SU(2), the holonomy transforms as
\begin{equation}\label{holGT}
h^{g}_\gamma  = g_{s(\gamma)} \, h_\gamma \, g^{-1}_{t(\gamma)},
\end{equation}
where $s(\gamma)$ and $t(\gamma)$ are respectively the source and target points of the line $\gamma$.
\item Under the action of diffeomorphism, the holonomy transforms as
\begin{equation}\label{holdiffeoaction}
h_\gamma\left( \phi^*A\right) = h_{\phi\circ\gamma}\left( A\right). 
\end{equation}
\item The functional derivative with respect to the connection gives 
\begin{equation}
\label{diffderholonomy}
\frac{\delta h_\gamma [A]}{\delta A^i_a(x)} = \left\lbrace\begin{array}{ll}
\f12\dot x^a \d^{(3)}(\gamma(s),x) \, \tau_i h_\gamma    & \text{if $x$ is the source of }\gamma  \\ 
\f12\dot x^a \d^{(3)}(\gamma(s),x) \, h_\gamma \tau_i   & \text{if $x$ is the target of }\gamma \\
\dot x^a \d^{(3)}(\gamma(s),x) \, h_\gamma(0,s) \tau_i h_\gamma(s,1)  & \text{if $x$ is inside } \gamma
\end{array}  \right. 
\end{equation}
\end{itemize}

\begin{computation}{Basic proofs}
If not otherwise specified, we assume for simplicity that the source is $t_0=0$.
For two composable paths $\alpha$  and $\beta$, we define the composition 
\begin{equation}\label{defComp}
\beta\alpha = \left\lbrace\begin{array}{c}
\beta(t), \qquad {\rm if} \ t \in [0,S]\\
\alpha(t-S), \qquad {\rm if} \ t \in [S,T+S] 

\end{array}  \right. 
\end{equation}
If we split the integrals as $\int_0^{S+T} = \int_0^{S} + \int_S^{S+T}$, \Ref{intmult} reads
\begin{equation}
\iiint_\square = \sum_{i=0}^n\binom{n}{i}\int_0^Sdt_1\cdots \int_0^Sdt_i\int_0^{T}dt_{i+1}\cdots\int_0^{T}dt_{n}.
\end{equation}
This allows us to split the path ordering as follows,
\begin{equation}
\mathcal P \left( A(\beta\alpha(t_1)) \cdots A(\beta\alpha(t_n))\right) = \mathcal P \left( A(\beta(t_{i+1})) \cdots A(\beta(t_n))\right)  \mathcal P \left( A(\alpha(t_1)) \cdots A(\alpha(t_i))\right).
\end{equation}
Hence, the holonomy along the composite path is 
\begin{align*}
h_{\beta\alpha} &= \sum_{n = 0}^{\infty} \sum_{i=0}^{n} \f1{i!\,(n-i)!} 
\int_{\square(T)}\mathcal P \left( A(\beta(t_{i+1})) \cdots A(\beta(t_n))\right) \int_{\square(S)} \mathcal P \left( A(\alpha(t_1)) \cdots A(\alpha(t_i))\right) \\
&= \sum_{i,q=0}^{\infty}\f1{q!} 
\int_{\square(T)}\mathcal P \left( A(\beta(t_{1})) \cdots A(\beta(t_q))\right) \frac{(-1)^{i}}{i!}\int_{\square(S)}\mathcal P \left( A(\alpha(t_{1})) \cdots A(\alpha(t_i))\right)\\
&= h_\beta h_\alpha,
\end{align*}
which proves \Ref{prima}. In the second step, we defined $q=n-i$.

To prove \Ref{holGT}, let us introduce families of vectors $u(t)$ and $w(t)$ such that
\begin{equation}\label{pippo}
w(t) = u(t)g(\gamma(t))  =    h_\gamma(t)g(\gamma(t))u(0).
\end{equation}
$w(t)$ satisfies the following differential equation,
\begin{align*}
\dfrac{d}{dt}w(t) &=  h \dot g u(0) + \dot h g u(0) =  hg g^{-1} \dot g u(0) + h g g^{-1}  A g u(0)\\
& =h g \left( g^{-1} \dot g  + g^{-1}Ag\right) u(0) = \left( g \dot g^{-1} + g^{-1} Ag\right) ghu(0) =  A^g w(t),
\end{align*}
which implies that $w(t)=h^g_\gamma(t) w(0) = h^g_\gamma(t) g(\gamma(0)) u(0) $. Comparing with \Ref{pippo} we find \Ref{holGT}.

The action of a diffeomorphism $\phi$ on the line integral of the connection is given by
\begin{equation}
\int_\gamma \phi A =\int_0^1A_\mu\left(\phi \circ \gamma(t)\right) \partial_\nu \phi^\mu\left( \gamma(t)\right) \dot \gamma^\nu(t) dt = \int_0^1 A_\mu\left(\phi \circ \gamma(t)\right)  \dfrac{d}{dt}\left(\phi \circ \gamma\right)^\mu(t) dt,
\end{equation}
which implies immediately 
\begin{equation}
h_\gamma\left( \phi A\right) = h_{\phi\circ\gamma}\left( A\right) .
\end{equation}

Finally, to prove \Ref{diffderholonomy}, let us compute the differential equation satisfied by $\frac{\delta h_\gamma(t_0,t)}{\delta A^i_a(x)}$ in the two cases of $x$ being inside the path or at one boundary. If $x$ is inside $\gamma$, 
from \eqref{holonomyode} we get
\begin{equation}
\frac{\mathrm d}{\mathrm d t} \frac{\delta h_\gamma(t_0,t)}{\delta A^i_a(x)} - \frac{\delta h_\gamma(t_0,t)}{\delta A^i_a(x)}A(\gamma(t)) = 0,
\end{equation}
which is solved by $\frac{\delta h_\gamma(t_0,t)}{\delta A^i_a(x)} =  h(t_0,s) \tau_i \dot x^a h(s,t)$. 
Then, by the Leibniz rule we also know that
\begin{equation}
\frac{\delta h_\gamma(t_0,t)}{\delta A^i_a(x)} = \frac{\delta h_\gamma(t_0,s)}{\delta A^i_a(x)}h_\gamma(s,t) + h_\gamma(t_0,s)\frac{\delta h_\gamma(s,t)}{\delta A^i_a(x)} =  h(t_0,s) \tau_i \dot x^a h(s,t).
\end{equation}
From this we can argue that if $x$ is at one boundary we must have
\begin{equation}
\frac{\delta h_\gamma(t_0,t)}{\delta A^i_a(x)} = \left\lbrace\begin{array}{ll}
\frac{1}{2}\dot x^a \tau_i h_\gamma(t_0,t)   & \text{if $x$ is the source of }\gamma \\ 
\frac{1}{2}\dot x^a h_\gamma(t_0,t)\tau_i   & \text{if $x$ is the target of }\gamma
\end{array}  \right. 
\end{equation}
\end{computation}

\subsection{Summary}

In this section we have taken the two key steps needed to prepare general relativity for the loop quantization.
The first step was to reformulate the theory in terms of the tetrad field and an SU(2) independent connection, the Ashtekar-Barbero connection \Ref{defA}. The second step was to regularize the resulting Poisson algebra using paths and surfaces, instead of the all of space as in traditional smearings (cf. what was done in the ADM formulation).
The resulting smeared algebra of $h_\gamma[A]$ and $E_i(S)$ is called \emph{holonomy-flux algebra}. It provides the most natural regular (i.e. no delta functions appear) version of the Poisson algebra \Ref{AEalgebra}.

\section{Loop quantum gravity: Kinematics}\label{SecKin}

The formulation of General Relativity in terms of the Ashtekar-Barbero connection and the densitized triad let us to talk about General Relativity in the language of a $SU(2)$ gauge theory, with Poisson brackets \Ref{AEalgebra} and the three sets of constraints
\begin{align*}
&G_i=0& &\text{Gauss law}\\
&H_a =0& & \text{Spatial diffeomorphism invariance}\\
&H =0& & \text{Hamiltonian constraint}
\end{align*}
The difference with a gauge theory is of course in the dynamics: In gauge theory, after imposing the Gauss law, we have a physical Hamiltonian. Here instead we still have a fully constrained system. 

Let us first briefly review some basic steps in the quantization of gauge theories.
In a nutshell, the usual procedure goes along these lines:
\begin{itemize}
\item Use the Minkowski metric to define a Gaussian measure $\delta A$ on the space of connections modulo gauge-transformations.
\item Consider the Hilbert space $L_2\left(A,\delta A\right)\ni \psi [A]$ and define the Schr\"odinger representation
\begin{subequations}\label{SchroYM}\begin{align}
\hat A^i_a \psi[A] &=  A^i_a \psi[A],\\
\hat E_i^a \psi[A] &= - i \hbar \gamma \frac{\delta}{\delta A^i_a} \psi[A],
\end{align}\end{subequations}
which satisfies the canonical commutation relation,
\begin{equation}
\left[ \hat A^i_a(x), \hat E^b_j(y) \right] = i \hbar \gamma \delta^b_a \delta^i_j \delta^3(x,y).
\end{equation}
\item Impose the Gauss law constraint, $\hat G_i \psi[A]=0$, which selects the gauge-invariant states. 
\item Study the dynamics with the physical Hamiltonian, 
\begin{equation}
i \partial_t \psi[A] = \hat H \psi[A].
\end{equation}
In practice, the easiest way to deal with the dynamics is to decompose $A$ in plane waves and the Hilbert space into a Fock space, $L_2(A,\delta A) = \HH_{Fock} = \bigoplus_{n}\HH_n$, the direct sum over $n$-particles spaces.
\end{itemize}

\medskip

The key difference in the case of general relativity is that we do not have a background metric at disposal to define the integration measure, since now the metric is a fully dynamical quantity.
Hence, we need to define a measure on the space of connections without relying on any fixed background metric.
The key to do this is the notion of cylindrical functions, which we introduce next.

\begin{computation}{Commutator}
\begin{align*}
\left[ \hat A^i_a(x), \hat E^b_j(y) \right] \psi[A] &=  - i \hbar \gamma A^i_a(x)  \frac{\delta}{\delta A^j_b(y)} \psi[A] +  i \hbar \gamma  \frac{\delta}{\delta A^j_b(y)}  A^i_a(x) \psi[A] \\
&= i \hbar \gamma \frac{\delta A^i_a(x)}{\delta A^j_b(y)}   \psi[A] = i \hbar \gamma \delta^b_a \delta^i_j \delta^3(x,y) \psi[A]
\end{align*}
\end{computation}

\subsubsection{Cylindrical functions and the kinematical Hilbert space}
Roughly speaking, a cylindrical function is a functional of a field that depends only on some subset of components of the field itself. In the case at hand, the field is the connection, and the cylindrical functions are functionals that depend on the connection only through the holonomies $h_e[A] = \mathcal{P}\exp\left(\int_e A\right)$ along some finite set of paths $e$. Consider a graph $\Gamma$, defined as a collection of oriented paths $e \subset \Sigma$ (we will call these paths the \emph{links} of the graph) meeting at most at their endpoints. Given a graph $\Gamma \subset \Sigma$ we denote by $L$ the total number of links that it contains.
A cylindrical function is  a couple $(\Gamma, f)$ of a graph and a a smooth function $f\,:\, SU(2)^{L} \longrightarrow \mathbb{C}$, and it is given by a functional of the connection defined as
\begin{equation}
\left\langle A \mid \Gamma,f \right\rangle = \psi_{(\Gamma,f)}[A] = f(h_{e_1}[A],\ldots,h_{e_{L}}[A]) \in Cyl_\Gamma
\end{equation}
where $e_i$ with $i=1, \ldots,L$ are the links of the corresponding graph $\Gamma$. 
\begin{center}
\includegraphics[width=10cm]{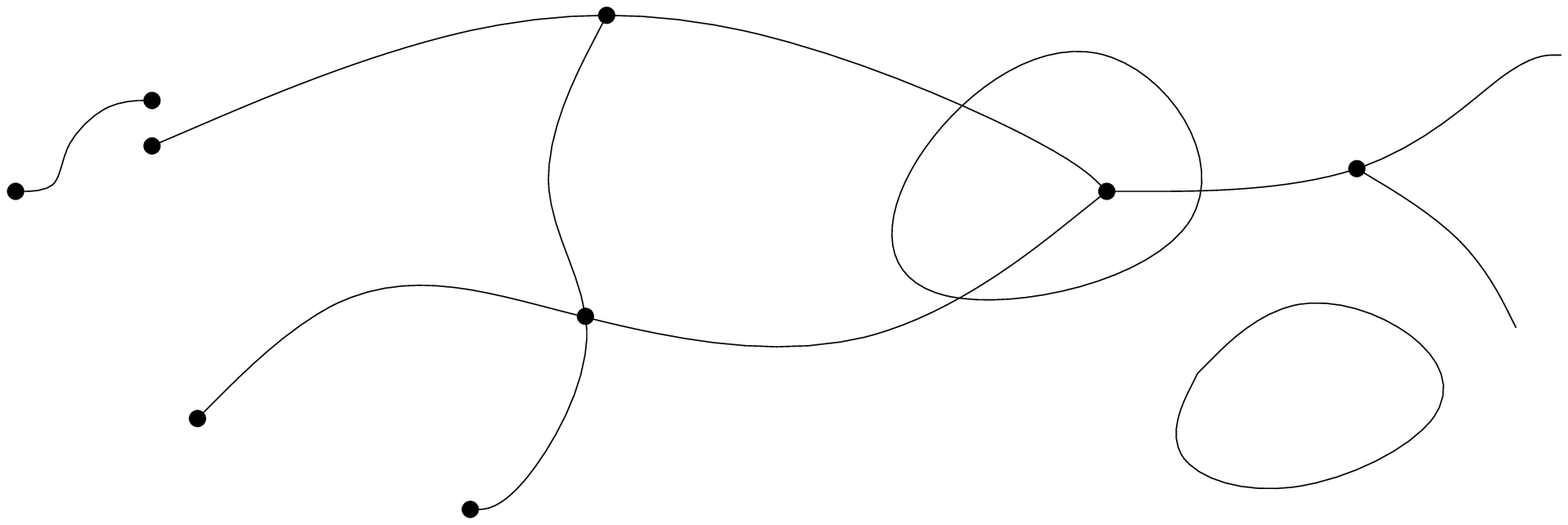} \\Collection of paths $\Gamma=\left\lbrace e_1,\ldots,e_n\right\rbrace$
\end{center}

This space of functionals can be turned into an Hilbert space if we equip it with a scalar product. The switch from the connection to the holonomy  is crucial in this respect, because the holonomy is an element of SU(2), and the integration over SU(2) is well-defined. In particular, there is a unique gauge-invariant and normalized measure d$h$, called the Haar measure. Using $L$ copies of the Haar measure, we define on $Cyl_\Gamma$ the following scalar product,
\begin{equation}
\label{scalarproductfixedgraph}
\left\langle \psi_{(\Gamma,f)} \mid \psi_{(\Gamma,f')} \right\rangle \equiv \int \prod_e \mathrm{d}h_e \overline{ f(h_{e_1}[A],\ldots,h_{e_L}[A])}  f'(h_{e_1}[A],\ldots,h_{e_L}[A]).
\end{equation}
This turns $Cyl_\Gamma$ into an Hilbert space $\HH_\Gamma$ associated to a given graph $\Gamma$.

Next, we define the Hilbert space of all cylindrical functions for all graphs as the direct sum of Hilbert spaces on a given graph,
\begin{equation}\label{Hkin}
\HH_{kin} = \underset{\Gamma \subset \Sigma}{\oplus} \HH_\Gamma.
\end{equation}
The scalar product on $\HH_{kin}$ is easily induced from \eqref{scalarproductfixedgraph} in the following manner: if $\psi$ and $\psi'$ 
share the same graph, then \eqref{scalarproductfixedgraph} immediately applies. If they have different graphs, say $\Gamma_1$ and $\Gamma_2$, we consider a further graph $\Gamma_3\equiv \Gamma_1 \cup \Gamma_2$, we extend $f_1$ and $f_2$ trivially on $\Gamma_3$, and define the scalar product as \eqref{scalarproductfixedgraph} on $\Gamma_3$:
\begin{equation}
\label{KinScalarProduct}
\left\langle \psi_{(\Gamma_1,f_1)} \mid \psi_{(\Gamma_2,f_2)} \right\rangle \equiv \left\langle \psi_{(\Gamma_1 \cup \Gamma_2,f_1)} \mid \psi_{(\Gamma_1 \cup \Gamma_2,f_2)} \right\rangle.
\end{equation}
The key result, due to Ashtekar and Lewandowski, is that \Ref{Hkin} defines an Hilbert space over (suitably generalized, see \cite{Ashtekar} for details) gauge connections $A$ on $\Sigma$, i.e.

\be\label{HkinL2}
\HH_{kin} = L_2[A, {\rm d}\mu_{AL}] .
\ee
The integration measure ${\rm d}\mu_{AL}$ over the space of connections is called the Ashtekar-Lewandowski measure. What \Ref{HkinL2} means is that \Ref{KinScalarProduct} can be seen as a scalar product between cylindrical functionals of the connection with respect to the Ashtekar-Lewandowski measure:
\be
\int \mathrm{d}\mu_{AL} \, \overline{\psi_{(\Gamma_1,f_1)}(A)} \psi_{(\Gamma_2,f_2)}(A) \equiv 
\left\langle \psi_{(\Gamma_1,f_1)} \mid \psi_{(\Gamma_2,f_2)} \right\rangle .
\ee

Now that we have a candidate kinematical Hilbert space which does not require a background metric, let us look for a representation of the holonomy-flux algebra on it. To that end, it is convenient to introduce an orthogonal basis in the space.
This can be done easily thanks to the Peter-Weyl theorem. It states that a basis on the Hilbert space $L_2(G,\mathrm{d}\mu_{Haar})$ of functions on a compact group $G$ is given by the matrix elements of the unitary irreducible representation of the group.
For the case of $SU(2)$, 
\begin{equation}
f(g)= \sum_j \hat f^j_{mn} D^{(j)}_{mn}(g)\qquad \begin{array}{c}j=0,\frac{1}{2},1,\ldots \\m=-j,\ldots, j\end{array} 
\end{equation}
where the Wigner matrices $D^{(j)}_{mn}(g)$ give the spin-j irreducible matrix representation of the group element $g$.
This immediately applies to $\HH_\Gamma$, since the latter is just a tensor product of $L_2(SU(2),\mathrm{d}\mu_{Haar})$. That is, the basis elements are
\begin{equation}
\label{PWbasis}
\left\langle A \mid \Gamma; j_e,m_e,n_e\right\rangle \equiv
D^{(j_1)}_{m_1n_1}(h_{e_1})\ldots D^{(j_n)}_{m_nn_n}(h_{e_n}),
\end{equation}
and a function $\psi_{(\Gamma,f)}[A] \in \HH_\Gamma$ can be decomposed as
\begin{equation}\label{fexp}
\psi_{(\Gamma,f)}[A]   =  \sum_{j_e,m_e,n_e} \hat f^{j_1,\ldots,j_n}_{m_1,\ldots,m_n,n_1,\ldots,n_n} D^{(j_1)}_{m_1n_1}(h_{e_1}[A])\ldots D^{(j_n)}_{m_nn_n}(h_{e_n}[A]).
\end{equation}

On this basis, we can give a Schr\"odinger representation like \Ref{SchroYM} for the regularized holonomy-flux version of the algebra. Consider for simplicity the fundamental representation, $h_e\equiv D^{(\f12)}(h_e)$. The holonomy acts by multiplication, 
\begin{subequations}\label{HF}\begin{equation}\label{holaction}
\hat h_\gamma[A] h_e[A] = h_\gamma[A] h_e[A],
\end{equation}
and the flux through the derivative \Ref{diffderholonomy},
\begin{equation}
\label{fluxaction}
\hat E_i(S) h_e[A] = 
- i \hbar \gamma \, \int_S\mathrm{d}^2 \sigma n_a \f{\d h_e[A]}{\d A^i_a(x(\sigma))}  = \pm i \hbar \gamma h_{e_1} [A] \tau_i h_{e_2}[A].
\end{equation}\end{subequations}
Here $e_1$ and $e_2$ are the two new edges defined by the point at which the triad acts and the sign depends on the relative orientation of $e$ and $S$. The action vanishes, $\hat E[S]h_{e_1}[A] = 0$, when $e$ is tangential to $S$ or $e \cap S = 0$.

\begin{computation}{Action of the Flux}
In the case $e\cup S = {p}$ recalling \eqref{diffderholonomy} we have
\begin{align}
\hat E_i(S) h_e =& - i \hbar \gamma \int_S\mathrm{d}^2 \sigma n_a \int_0^1 \mathrm{d}s \dot{x}^a(s) \delta^3 (p,x(\sigma)) h_{e_1}\tau_i h_{e_2} =\\
& - i \hbar \gamma(-1)^{o(p)}h_{e}(0,s)\tau_i h_e(s,1)
\end{align}
where $(-1)^{o(p)}= \pm 1$ depending if $\sigma_1$, $\sigma_2$, $s$ is a right or left handed coordinates system
\begin{equation}
\int_S\int_0^1 \mathrm{d}\sigma_1 \mathrm{d}\sigma_2\mathrm{d}s \varepsilon_{abc}\frac{\partial x^a}{\partial \sigma_1}\frac{\partial x^b}{\partial \sigma_2}\frac{\partial x^c}{\partial s} \delta^3 (x(s),x(\sigma)) = \pm \int \mathrm{d}^3 x \delta(x) = \pm 1
\end{equation}
If there are no intersection the integration of the $\delta^3 (p,x(\sigma))$ is $0$. In the case of $e$ tangent to $S$ we can compute the action of $E$ as a limit of a double intersection. Because the two contributes have relative opposite sign the limit is trivially $0$.

\end{computation}
Consider now the action of the scalar product of two fluxes acting inside the link,
\be\label{EE1}
\hat E_i(S) \hat E^i(S) h_e[A]  = - \hbar^2 \gamma^2 \, h_{e_1} [A] \tau^i \tau_i h_{e_2}[A].
\end{equation}
On the right hand side, we see the appearance of the scalar contraction of algebra generators, $\tau^i \tau_i \equiv C^2$. This scalar product is known as the Casimir operator of the algebra. In the fundamental representation considered here, $C^2=-\f34 {\mathbbm 1}_2$. The Casimir clearly commutes with all group elements, thus \Ref{EE1} can be written as
\be\label{EE2}
\hat E_i(S) \hat E^i(S) h_e[A]  = - \hbar^2 C^2 \gamma^2 \, h_{e_1} [A] h_{e_2}[A] = - \hbar^2 C^2 \gamma h_{e} [A].
\end{equation}
This expression will be useful below.

On the other hand, if two consecutive fluxes act on one endpoint, say the target, we get
\be\label{Eq}
\hat E_i(S) \hat E_j(S) h_e[A]  = - \hbar^2 \gamma^2 \, h_{e} [A] \tau_i \tau_j.
\end{equation}
From this result we immediately find that two flux operators \emph{do not commute},
\be\label{Enc}
[\hat E_i(S), \hat E_j(S)] h_e[A]  = - \hbar^2 \gamma^2 \, h_{e} [A] [\tau_i, \tau_j]
= - \hbar^2 \gamma^2 \epsilon_{ij}{}^{k} \, h_{e} [A] \tau_k.
\end{equation}

The actions \Ref{HF} of the holonomy-flux algebra trivially extends to a generic basis element $D^{(j)}(h)$. The action \Ref{holaction} is unchanged, and in the right hand side of \Ref{fluxaction} one simply has to replace $\tau_i$ by the generator $J_i$ in the arbitrary irreducible $j$. Consequently, in \Ref{EE2} we have the Casimir $C^2_j=-j(j+1) {\mathbbm 1}_{2j+1}$ on a generic irreducible representation,
\be\label{EE3}
\hat E_i(S) \hat E^i(S) D^{(j)}(h_e)   = \hbar^2 \gamma^2 j(j+1) \, D^{(j)}(h_e).
\end{equation}

Finally, the action is extended by linearity over the whole $\HH_{kin}$. The remarkable fact is that this representation of the holonomy-flux algebra on $\mathcal H_{kin}$ is \emph{unique}, as proved by Fleischhack and Lewandowski, Okolow, Sahlmann, Thiemann \cite{Lewandowski:2005jk}. 
This uniqueness result can be compared to the Von Neumann theorem in quantum mechanics on the uniqueness of the Schr\"odinger representation. It is well-known that the uniqueness does not extend to interacting field theories on flat spacetime. Remarkably, insisting on background-independence reintroduces such uniqueness also for a field theory.

What we have accomplished with this construction is the definition of a well-behaved kinematical Hilbert space for general relativity.
It carries a representation of the canonical Poisson algebra, and as a bonus, this representation is unique. Following Dirac, we now have a well-posed problem of reduction by the constraints:
\be \label{Dirac1}
\HH_{kin} \ \ \xrightarrow{\displaystyle \hat G_i=0} \ \ \HH^0_{kin} \ \ \xrightarrow{\displaystyle \hat H^a=0} \ \  \HH_{Diff} \ \ \xrightarrow{\displaystyle \hat H=0\phantom{^a} } \ \ \HH_{phys}.
\ee

\subsection{Gauge-invariant Hilbert space}

The first step is to find the solutions of the quantum Gauss constraint. These are the states in $\HH_{kin}$ that are $SU(2)$ gauge invariant. These solutions define a new Hilbert space, that we call $\HH_{kin}^0$ where we leave the subindex $kin$ to keep in mind that there are still constraints to be solved before arriving to $\HH_{phys}$. 
The action of the Gauss constraint is easily represented in $\HH_{kin}$. In fact, recall that under gauge transformations
\begin{equation}
h_e \longrightarrow h_e ' = \hat U_G h_e = g_{s(e)} h_e g^{-1}_{t(e)}.
\end{equation}
Similarly, in a generic irrep $j$ we have 
\begin{equation}
D^{(j)}(h_e) \longrightarrow D^{(j)}(h_e ') = D^{(j)}(g_{s(e)} h_e g^{-1}_{t(e)}) = D^{(j)}(g_{s(e)}) D^{(j)}(h_e) D^{(j)}(g^{-1}_{t(e)}).
\end{equation}
From this it follows that gauge transformations act on the source and targets of the links, namely on the \emph{nodes} of a graph. Imposing gauge-invariance then means requiring the cylindrical function to be invariant under action of the group at the nodes:
\be\label{gif0}
f_0(h_1, \ldots , h_L) \equiv f_0(g_{s_1} h_1 g_{t_1}{}^{-1}, \ldots, g_{s_L} h_L g_{t_L}{}^{-1})
\ee
This property can be easily implemented via \emph{group averaging}: given an arbitrary $f\in Cyl_\Gamma$, the function
\be\label{gaugeProj}
f_{0} (h_1, \ldots , h_L) \equiv \int \prod_n \mathrm{d}g_n \; f(g_{s_1} h_1 g_{t_1}{}^{-1}, \ldots , g_{s_L} h_L g_{t_L}{}^{-1})
\ee
clearly satisfies \Ref{gif0}.

%
%

\begin{computation}{Example: The theta graph} To give a constructive example, let us consider the following graph,\\
\begin{center}
\includegraphics[scale=0.8]{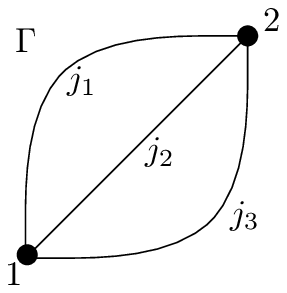} \\
\end{center}
A generic cylindrical function can be expressed in terms of the orthonormal basis using \Ref{fexp}. Since the gauge transformations act only on the group elements, the gauge-invariant part is obtained looking at the gauge-invariant part of the product of Wigner matrices, 
\begin{equation}\nonumber
f_{inv}(h_1, \ldots, h_3) = \sum_{j_l,m_l,n_l} \hat f^{j_1,j_2,j_3}_{m_1,m_2,m_3,n_1,n_2,n_3} \left[ D^{(j_1)}_{m_1n_1}(h_{1}) D^{(j_2)}_{m_2n_2}(h_{2}) D^{(j_3)}_{m_3n_3}(h_{3})\right]_{inv}.
\end{equation}
Using the definition \Ref{gaugeProj}, the invariant part of the basis is
\begin{align}\nonumber
&\left[ D^{(j_1)}_{m_1n_1}(h_{1}) D^{(j_2)}_{m_2n_2}(h_{2}) D^{(j_3)}_{m_3n_3}(h_{3})\right]_{inv} =\\\nonumber
& = \int \mathrm{d}g_1\mathrm{d}g_2 D^{(j_1)}_{m_1n_1}(g_1h_{1}g_2^{-1}) D^{(j_2)}_{m_2n_2}(g_1h_{2}g_2^{-1}) D^{(j_3)}_{m_3n_3}(g_1h_{3}g_2^{-1})=\\\nonumber
& =\mathcal{P}_{m_1 m_2 m_3 \alpha_1 \alpha_2 \alpha_3} \mathcal{P}_{\beta_1 \beta_2 \beta_3 n_1 n_2 n_3} D^{(j_1)}_{\alpha_1\beta_1}(h_{1}) D^{(j_2)}_{\alpha_2\beta_2}(h_{2}) D^{(j_3)}_{\alpha_3\beta_3}(h_{3}),
\end{align}
where $\mathcal{P}_{m_1 m_2 m_3 \alpha_1 \alpha_2 \alpha_3}$ is the projector on the gauge invariant space,
\begin{equation}\nonumber
\mathcal{P}_{m_1 m_2 m_3 \alpha_1 \alpha_2 \alpha_3} = \int \mathrm{d}g_1 D^{(j_1)}_{m_1\alpha_1}(g_1) D^{(j_2)}_{m_2\alpha_2}(g_1) D^{(j_3)}_{m_3\alpha_3}(g_1).
\end{equation}
This projector can be written in terms of normalized Clebsch-Gordan coefficients, or Wigner's 3j-m symbols, as
\begin{equation}\label{3jm}
\int \mathrm{d}g_1 D^{(j_1)}_{m_1\alpha_1}(g_1) D^{(j_2)}_{m_2\alpha_2}(g_1) D^{(j_3)}_{m_3\alpha_3}(g_1) = \begin{pmatrix}
  j_1 & j_2 & j_3\\
  m_1 & m_2 & m_3
\end{pmatrix}\overline{\begin{pmatrix}
  j_1 & j_2 & j_3\\
  \alpha_1 & \alpha_2 & \alpha_3
\end{pmatrix}}.
\end{equation}
With this notation,
\begin{align}\nonumber
&\left[ D^{(j_1)}_{m_1n_1}(h_{1}) D^{(j_2)}_{m_2n_2}(h_{2}) D^{(j_3)}_{m_3n_3}(h_{3})\right]_{inv} =\\\nonumber
&=\begin{pmatrix}
  j_1 & j_2 & j_3\\
  m_1 & m_2 & m_3
\end{pmatrix}\overline{\begin{pmatrix}
  j_1 & j_2 & j_3\\
  \alpha_1 & \alpha_2 & \alpha_3
\end{pmatrix}}\begin{pmatrix}
  j_1 & j_2 & j_3\\
  \beta_1 & \beta_2 & \beta_3
\end{pmatrix}
\overline{\begin{pmatrix}
  j_1 & j_2 & j_3\\
  n_1 & n_2 & n_3
\end{pmatrix}}D^{(j_1)}_{\alpha_1\beta_1}(h_{1}) D^{(j_2)}_{\alpha_2\beta_2}(h_{2}) D^{(j_3)}_{\alpha_3\beta_3}(h_{3})\\\nonumber
&=\begin{pmatrix}
  j_1 & j_2 & j_3\\
  m_1 & m_2 & m_3
\end{pmatrix}
\overline{\begin{pmatrix}
  j_1 & j_2 & j_3\\
  n_1 & n_2 & n_3
\end{pmatrix}}\prod_e D^{(j_e)}(h_{e}) \prod_n i_n
\end{align}
where $i_n$ is a short-hand notation for the 3j-m symbols. Notice that these are the invariant tensors in the space of $\underset{e \in n}{\otimes} j_e$ of all the spins that enters in the node $n$. Finally, we have
\begin{align}\nonumber
f_{inv} &= \sum_{j_e} \prod_e D^{(j_e)}(h_{e}) \prod_n i_n \sum_{m_e n_e }\hat f^{j_1,j_2,j_3}_{m_1,m_2,m_3,n_1,n_2,n_3} \begin{pmatrix}
  j_1 & j_2 & j_3\\
  m_1 & m_2 & m_3
\end{pmatrix}
\overline{\begin{pmatrix}
  j_1 & j_2 & j_3\\
  n_1 & n_2 & n_3
\end{pmatrix}} \\
&= \sum_{j_e} \hat f^{j_1,j_2,j_3}\prod_e D^{(j_e)}(h_{e}) \prod_n i_n,\label{example}
\end{align}
with the new coefficients $\hat f^{j_1,j_2,j_3}$ including the sums over the magnetic numbers $m_e$, $n_e$.

\end{computation}

The group averaging amounts to inserting on each node $n$ the following projector,
\begin{equation}\label{defP}
\mathcal P = \int \mathrm{d}g \prod_{e \in n} D^{(j_e)}(g).
\end{equation}
Here the integrand is an element in the tensor product of $SU(2)$ irreducible representations,
\begin{equation}
\prod_e D^{(j_e)}_{m_e n_e}(h_e) \in \bigotimes_e V^{(j_e)}.
\end{equation}
As such, it transforms non-trivially under gauge transformation and is in general reducible,
\begin{equation}
\bigotimes_e V^{(j_e)} = \bigoplus_i V^{(j_i)}.
\end{equation}
Then, the integration in \Ref{defP} selects the gauge invariant part of $\bigotimes_e V^{(j_e)}$, namely the \emph{singlet} space $V^{(0)}$, if the latter exists. Since $\cal{P}$ is a projector, we can decompose it in terms of a basis of $V^{(0)}$. Denoting $i_\alpha$ a vector (``ket'') in this basis, $\alpha=1, \ldots, {\rm dim} V^{(0)}$, and $ i_\alpha^*$ the dual (``bra''),
\begin{equation}
\label{projectorintertwiner}
{\cal P} = \sum_{\alpha = 1}^{{\rm dim} V^{(0)}} i_\alpha \, i_\alpha^*.
\end{equation}
These invariants are called \emph{intertwiners}.
For the case of a 3-valent node as in the above example, ${\rm dim} V^{(0)}=1$ and the unique intertwiner $i$ is given by Wigner's 3j-m symbols  (cf. \Ref{3jm}). More precisely in the case of a three-valent node the space \begin{equation*}
\left[ V^{(j_1)} \otimes V^{(j_2)} \otimes V^{(j_3)}\right]_{inv}
\end{equation*} is non-empty only when the following Clebsch-Gordan conditions hold, 
\begin{equation}
|j_2-j_3| \leq j_1 \leq j_2 + j_3.
\end{equation}
For an $n$-valent node, the space $V^{(0)}$ can have a larger dimension. To visualize the intertwiners, it is convenient to add first two irreps only, then the third, and so on. This gives rise to a decomposition over virtual links, which for $n=4$ and $n=5$ looks as follows:
\begin{center}
\includegraphics[scale=0.8]{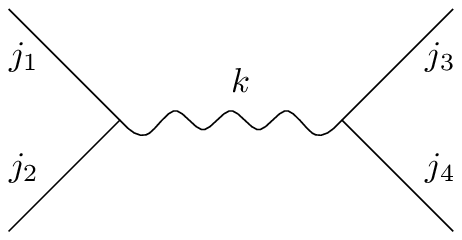} \hspace{1cm}
\includegraphics[scale=0.8]{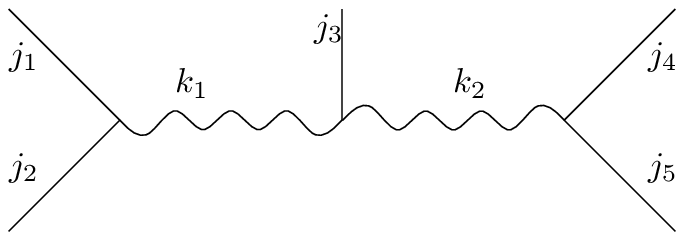} 
\end{center}
The virtual spins $k_i$ label the intertwiners.

\begin{computation}{Brief proof that $\mathcal P$ is a projector}
First we check that $\mathcal P \mathcal P = \mathcal P$
\begin{align}
\mathcal P\mathcal P &= \int \mathrm{d}g_1 \mathrm{d}g_2 \prod_{e\in n} D^{(j_e)}(g_1)\prod_{e\in n} D^{(j_e)}(g_2) = \int \mathrm{d}g_1 \mathrm{d}g_2 \prod_{e\in n} D^{(j_e)}(g_1) D^{(j_e)}(g_2) =\\
& =\int \mathrm{d}g_1 \mathrm{d}g_2 \prod_{e\in n} D^{(j_e)}(g_1 g_2) = \int \mathrm{d}g_2 \int  \mathrm{d}g_1 \prod_{e\in n} D^{(j_e)}(g_1) = \mathcal P 
\end{align}
than that is left and right invariant
\begin{equation}
\prod_{e\in n} D^{(j_e)}(g_1) \mathcal P = \int  \mathrm{d}g_2 \prod_{e\in n} D^{(j_e)}(g_1)D^{(j_e)}(g_2) = \int  \mathrm{d}g_2 \prod_{e\in n} D^{(j_e)}(g_1 g_2) = \mathcal{P}
\end{equation}
\begin{equation}
\mathcal P\prod_{e\in n} D^{(j_e)}(g_1)  = \int  \mathrm{d}g_2  \prod_{e\in n} D^{(j_e)}(g_2) D^{(j_e)}(g_1)= \int  \mathrm{d}g_2  \prod_{e\in n} D^{(j_e)}(g_2 g_1) = \mathcal{P}
\end{equation}
\end{computation}\\

The facts that $\mathcal{P}$ acts only on the nodes of the graph that label the basis of $\HH_{kin}$  and equation \eqref{projectorintertwiner} implies that the result of the action of $\mathcal{P}$ on elements of $\HH_{kin}$ can be written as a linear combination of products of representation matrices $D^{(j)}(h_e)$ contracted with intertwiners, generalizing the result \Ref{example}. The states labeled with a graph $\Gamma$, with an irreducible representation $D^{(j)}(h)$ of spin-j of the holonomy $h$ along each link, and with an element $i$ of the intertwiner space $\HH_n \equiv {\rm Inv}[\underset{e \in n}{\otimes} V^{(j_e)}]$ in each node, are called spin network states, and are given by
\begin{equation}
\label{spinnetworkbasis}
\psi_{(\Gamma, j_e, i_n)}[h_e] = \underset{e}{\otimes} D^{(j_e)}(h_e) \underset{n}{\otimes}i_n.
\end{equation}
Here the indices of the matrices and of the interwiners are hidden for simplicity of notation. Their contraction pattern can be easily reconstructed from the connectivity of the graph.
\begin{center}
\includegraphics[scale=0.8]{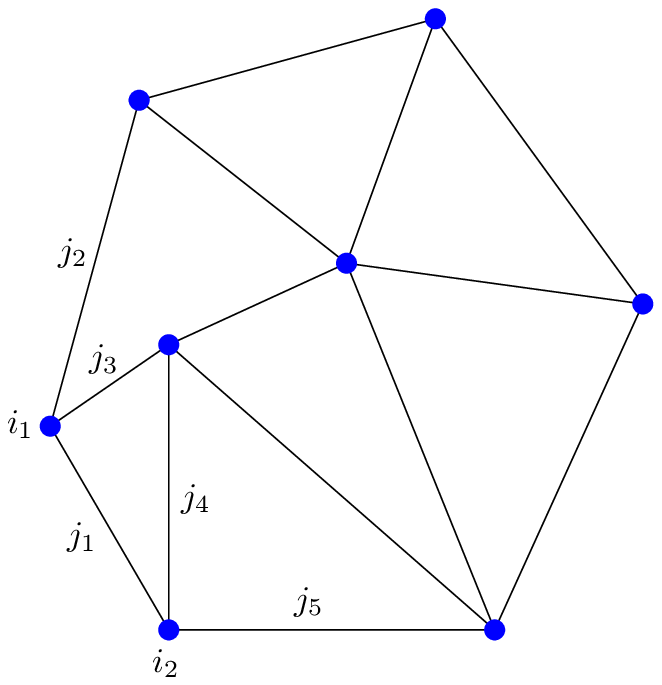}
\end{center}

To complete the discussion, let us show that imposing the gauge-invariance amounts to solving the Gauss law constraint, $\hat G_i \psi=0$. Consider a gauge invariant node $n$, and a surface $S$ centered in $n$ of radius $\epsilon$. The action of the total flux operator through $S$ on $n$ vanishes identically:
\begin{equation}
\label{gauss}
\lim_{\epsilon\to 0}\hat E(S) \left| n \right\rangle  = 0.
\end{equation}
In fact, using \eqref{fluxaction} at each link one notices that \eqref{gauss} produces the infinitesimal gauge transformation $g_\alpha = 1 - \alpha_i \tau^i \in SU(2)$ at the node, and because the node is gauge invariant such action vanishes. 

Summarizing, spin network states \Ref{spinnetworkbasis} 
form a complete basis of the Hilbert space of solutions of the quantum Gauss law, $\HH_{kin}^0$. The structure of this space is nicely organized by the spin networks basis. As before, different graphs $\Gamma$ select different orthogonal subspaces, thus $\HH_{kin}^0$ decomposes as a direct sum over spaces on a fixed graph,
\be\label{Hkin00}
\HH_{kin}^0 = \underset{\Gamma \subset \Sigma}{\oplus} \HH^0_\Gamma.
\ee
Furthermore, the Hilbert space on a fixed graph decomposes as a sum over intertwiner spaces,
\be\label{Hkin0}
\HH_{\Gamma}^0 = L_2[SU(2)^L/SU(2)^N, {\rm d}\mu_{Haar}]= \oplus_{j_l} \left(\otimes_n {\cal H}_{n} \right).
\ee
These two equations are the analogue in loop gravity of the Fock decomposition of the Hilbert space of a free field in Minkowski spacetime into a direct sum of $n$-particle states, and play an equally important fundamental role.

\subsection{Geometric operators}\label{SecGO}
\subsubsection{The area operator}

The simplest geometric operator that can be constructed in loop quantum gravity is the area operator.
The area of a surface $S$ can be given in terms of its normal $n_a$ and the densitized triad $E_i^a$,
\begin{equation}
A(S) = \int_S \mathrm{d}\sigma_1 \mathrm{d}\sigma_2 \sqrt{E^a_i E^{bi} n_a n_b}
\end{equation}

\begin{computation}{The area at classical level}
We start from the standard definition of area in terms of the metric,
\begin{equation}
A(S)  = \int_S \mathrm{d}\sigma_1 \mathrm{d}\sigma_2 \sqrt{\mathrm{det} \left( g_{ab}\frac{\partial x^a}{\partial \sigma^\alpha}\frac{\partial x^b}{\partial \sigma^\beta} \right) } \qquad \alpha,\beta = 1,2
\end{equation}
using the notation $\partial_1 x^a = \frac{\partial x^a}{\partial \sigma^1}$
\begin{align}
\mathrm{det} \left( g_{ab}\frac{\partial x^a}{\partial \sigma^\alpha}\frac{\partial x^b}{\partial \sigma^\beta} \right) &= g_{ab}g_{cd} \left[ \partial_1 x^a \partial_1 x^b \partial_2 x^c \partial_2 x^d -  \partial_1 x^a \partial_2 x^b \partial_1 x^c \partial_2 x^d \right] =\\
&= g_{ab}g_{cd} 2 \partial_1 x^a \partial_1 x^{[b} \partial_2 x^{c]} \partial_2 x^{d} = 2  g_{a[b}g_{c]d} \partial_1 x^a \partial_1 x^{b} \partial_2 x^{c} \partial_2 x^{d}=\\ &=g g^{ef}n_e n_f
\end{align}
where we recognized that
\begin{equation}
 g_{a[b}g_{c]d} = \frac{1}{2} \varepsilon_{ace}\varepsilon_{bdf} g g^{ef} \qquad n_e = \varepsilon_{eab} \frac{\partial x^a}{\partial \sigma^1}\frac{\partial x^b}{\partial \sigma^2}
\end{equation}
Using the definition of tetrad and densitized tetrad we have
\begin{equation}
A(S) =  \int_S \mathrm{d}\sigma_1 \mathrm{d}\sigma_2 \sqrt{ e^2 e^e_i e^{fi} n_e n_f} = \int_S \mathrm{d}\sigma_1 \mathrm{d}\sigma_2 \sqrt{ E^e_i E^{fi} n_e n_f} 
\end{equation}
\end{computation}

At the quantum level, we know from \Ref{fluxaction} that the triad operators act as functional derivatives. The action of the scalar product of two triads operator was also studied, see equation \Ref{EE3}, for the case of a surface intersected only once by the holonomy path. The case of a generic graph can be easily dealt with if we regularize the expression for the area in the following way. We introduce a decomposition of $S$ in $N$ two-dimensional cells, and write the integral as the limit of a Riemann sum, 
\begin{equation}
A(S) = \lim_{N\to \infty} A_N(S),
\end{equation}
where the Riemann sum can be expressed as
\begin{equation}
\label{classicalarea}
A_N(S) = \sum_{I=1}^N \sqrt{E_i(S_I) E^i(S_I)}.
\end{equation}
Here $N$ is the number of cells, and $E_i(S_I)$ is the flux of $E_i$ through the $I$-th cell.

\begin{computation}{Checking the limit}
In the limit  of infinitesimal cells we have that
\begin{equation}
E_i(S_I)\equiv \int_{S_I} n_a E^a_i \mathrm{d}^2 \sigma = \int_{S_I} E_i^a n_a \approx E_i^a n_a S_I
\end{equation}
In that limit the definition of the area
\begin{align}
A_N(S) &= \sum_{I=1}^N \sqrt{E_i(S_I) E^i(S_I)} \approx \sum_{I=1}^N \sqrt{E_i^a n^I_a S_IE^{bi} n^I_b S_I} = \sum_{I=1}^N S_I\sqrt{E_i^a n^I_a E^{bi} n^I_b} =\\
&=\int_S \sqrt{ E^a_i E^{bi} n_a n_b} 
\end{align}
\end{computation}

Accordingly, we define the area operator as
\begin{equation}
\hat{A}(S) = \lim_{N\to \infty} \hat A_N(S),
\end{equation}
where in $A_N(S)$ we simply replace the classical flux $E_i(S_I)$ by the operator $\hat E_i(S_I)$. This operator now acts on a generic spin network state $\psi_\Gamma$, where the graph $\Gamma$ is generic and can intersect $S$ many times. We already know that $\hat E_i(S_I) \hat E^i(S_I)$ gives zero if $S_I$ is not intersected by any link of the graph. Therefore once the decomposition is sufficiently fine so that each surface $S_I$ is punctured once and only once, taking a further refinement has no consequences. Therefore, the limit amounts to simply sum the contributions of the finite number of punctures $p$ of $S$ caused by the links of $\Gamma$. That is, 
\begin{equation}
\hat A(S) \, \psi_\Gamma = \lim_{N\to\infty} \sum_{I=1}^N \sqrt{\hat E_i(S_I) \hat E^i(S_I)} \, \psi_\Gamma = \sum_{p\in S\cup \Gamma} \hbar \sqrt{\gamma^2 j_p (j_p+1)} \, \psi_\Gamma.
\end{equation}

There are two key remarks to make to this formula: first of all, the spectrum of the area operator is completely known\footnote{
In this expression, we assumed that each puncture is caused by a link crossing the surface. However, it could also happen that $S$ is puncture by a node of the graph. A closed expression for the area spectrum is known also in this general case. See the literature for details \cite{RovelliBook, ThiemannBook, Ashtekar:1997fb}.} and \emph{quantized}: the area can only take up discrete values, with minimal excitation being proportional to the squared Planck length $\ell_P^2 =\hbar G$, restoring Newton's constant.
This result can be compared with other celebrated quantizations, such as the radii of electron's orbitals in atoms.

Second, the operator has a \emph{diagonal} action on spin networks. Therefore, spin network states are eigenstates of the area operator.

\subsubsection{The volume operator}

Given a region $R\subset \Sigma$ classically we can define its volume as
\begin{equation}
\label{ClassicalVolume}
V(R) = \int_R \mathrm{d}^3 x \sqrt{g} =  \int_R \mathrm{d}^3 x \sqrt{\left| \frac{1}{3!} \varepsilon_{abc}\varepsilon^{ijk} E^a_iE^b_jE^c_k \right|},
\end{equation}
where the quantity in absolute value can be recognized as the determinant of the densitised triad, det $E^a_i$.

Two distinct mathematically well-defined volume operators have been proposed in the literature. One is due to Rovelli and Smolin, and the other to Ashtekar and Lewandowski. We will refer to them as RS and AL respectively.
Both of them act non-trivially only at the nodes of a spin network state. 
Let us begin reviewing the construction by RS.

As we did for the area, we replace the integral over $R$ by the limit of a Riemann sum. Specifically, we consider a partition of the region in cubic cells $C_I$ so that $R \subset \cup_I C_I$, and the integral $\int_R \mathrm{d}^3 x$ can be approximated from above by the sum $\sum_I \mathrm{volume}(C_I)$.
This partition allows us to rewrite \Ref{ClassicalVolume} in terms of fluxes. In fact, consider the following integral,
\begin{equation*}
W_I \equiv \frac{1}{48} \int_{\partial C_I} \mathrm{d}^2 \sigma_1 \int_{\partial C_I} \mathrm{d}^2 \sigma_2 \int_{\partial C_I} \mathrm{d}^2 \sigma_3 \left|\varepsilon_{ijk}E^a_i(\sigma_1) n_a(\sigma_1)E^b_j(\sigma_2) n_b(\sigma_2)E^c_k(\sigma_3) n_c(\sigma_3) \right|.
\end{equation*}
In the continuum limit where we send the size of the cell $\eps\mapsto 0$ and we shrink the cell to a point $x$, we obtain 
\begin{equation*}
W_I = \f1{48}\eps^{abc} n_a n_b n_c \, {\rm det} \, E^a_i(x) \,\eps^6 \simeq {\rm det} \, E^a_i(x) \,\eps^6 \simeq {\rm volume}^2(C_I).
\end{equation*}
Hence, we have 
\be
V(R) = \lim_{\eps\mapsto 0} \sum_I \sqrt{W_I}.
\ee

For the sake of notation, let us subdivide each $\partial C_I$ into surfaces $S^\alpha$ such that $\partial C_I = \cup_{\alpha} S^\alpha_I$. Then, we can write $W_I$ as a sum of fluxes over three surfaces, and
\begin{equation}
V (R) = \lim_{\eps\mapsto 0} \sum_I
\sqrt{ \frac{1}{48} \sum_{\alpha,\beta,\gamma} \left|\varepsilon_{ijk}E_i(S^\alpha_I) E_j(S^\beta_I)E_k(S^\gamma_I). \right|}
\end{equation}
Finally, we can simply turn the classical fluxes to operators, 
\begin{equation}\label{V3}
\hat V (R) = \lim_{\eps\mapsto 0} \sum_I
\sqrt{\frac{1}{48} \sum_{\alpha,\beta,\gamma} \left|\varepsilon_{ijk} \hat E_i(S^\alpha_I) \hat E_j(S^\beta_I)\hat E_k(S^\gamma_I)  \right|}.
\end{equation}
This is the Rovelli-Smolin volume operator.

As for the area operator, one finds that there exists an ``optimal'' subdivision, after which the result stays unchanged with any further refinement, and so the limit can be safely taken. For the area, this consisted in the small surfaces being punctured only once at most. Something similar happens for the volume. The ``optimal'' partition is reached as follows.
The nodes of $\Gamma$ can fall only in the interior of cells, and a cell $C_I$ contains at most one node. In case the cell contains no node, then we assume that it contains at most one link. 
Moreover, we assume that the partition of the surfaces $\partial C_I$ in cells $S^\alpha_I$ is refined so that links of $\Gamma$ can intersect a cell $S^\alpha_I$ only in its interior and each cell $S^\alpha_I$ is punctured at most by one link.

This said, let us now study the action of the operator. The first thing to notice is that the presence of the epsilon tensor requires all three fluxes to be different: if two are the same, then their antisymmetric combination introduced by the epsilon vanishes. In particular, this means that the volume does not act on links, since if no node is present, two of the $S_I^\alpha$ have to be the same.
We thus obtain the important result that \emph{the volume operator acts only on nodes of the graph}.

Let us now focus on a single node, i.e. the $I$-th contribution to \Ref{V3}. We consider the cubic operator
\begin{equation}\label{defU}
\hat U = \frac{1}{48} \sum_{\alpha,\beta,\gamma} \left|\varepsilon_{ijk} \hat E_i(S^\alpha) \hat E_j(S^\beta)\hat E_k(S^\gamma) \right|.
\end{equation}
Let us restrict to gauge-invariant spin networks, where the Gauss law holds and each node is labeled by an intertwiner. First of all, it is immediate to see that the action of \Ref{defU} on a 3-valent node is zero.
In fact, the Gauss law tells us that the sum of the fluxes through a surface around a gauge invariant node is zero. In the 3-valent case, only three $S^\alpha$ give non-zero fluxes, thus
\begin{equation}
\left(  \hat E_i(S^\alpha) + \hat E_i(S^\beta) + \hat E_i(S^\gamma) \right)  \left|i \right\rangle = 0,
\ee
which implies
\be \hat E_i(S^\alpha)\left|i \right\rangle = - \left( \hat E_i(S^\beta) + \hat E_i(S^\gamma) \right)  \left|i \right\rangle.
\end{equation}
Using this result in \Ref{defU} we get zero because two identical fluxes always appear, 
\begin{equation}\nonumber
\varepsilon_{ijk} \hat E_i(S^\alpha) \hat E_j(S^\beta)\hat E_k(S^\gamma)\left|i \right\rangle =
- \varepsilon_{ijk} \left( \hat E_i(S^\beta) + \hat E_i(S^\gamma) \right) \hat E_j(S^\beta)\hat E_k(S^\gamma) \left|i \right\rangle = 0.
\end{equation}

Non-trivial contributions to the volume comes from nodes of valency 4 or higher. Notice that these are the cases for which the intertwiner is not unique, but a genuine independent quantum number. Thus the operator \Ref{defU} probes precisely the degrees of freedom hidden in the intertwiners.
Consider the 4-valent case. Of all the surface cells $S^\alpha$, only four are punctured by the links. Hence, we have only four contributions to the Gauss law, which we can then use to eliminate one flux in favor of the remaining three,
\begin{equation}\label{Gauss4}
\hat E_i(S^4) = - \hat E_i(S^1) -\hat E_i(S^2) -\hat E_i(S^3).
\end{equation}
The sum over the $S^\alpha$ in \Ref{defU} then reduces to the sole contributions from the four punctured surfaces. Using \Ref{Gauss4} these contributions are all equal. A simple combinatorial shows that there are 48 terms, all equal, thus
\begin{equation}
\hat U = \left|\varepsilon_{ijk} \hat E_i(S^1)  \hat E_j(S^2)\hat E_k(S^3)\right| = \hbar^3  \left|\gamma^3 \varepsilon_{ijk} J^1_i  J^2_j J^3_k\right|.
\end{equation}
In the last step we have used the result \Ref{fluxaction}, which gives the action of the fluxes in terms of the generators $\vec J^a$ in the spin $j_a$ representation. Notice that the orientation factors $\pm$ (between the edge and the surface) are irrelevant because of the modulus.

This is a well-defined cubic operator, whose action spectrum is again discrete, with minimal excitation proportional to $(\hbar G)^{3/2}$. The explicit formula is much more complicated than that of the area. We refer the interested reader to the literature \cite{De Pietri:1996pja}.

\medskip

The alternative proposal by Ashtekar and Lewandowski has the extra property of being sensitive to the differential structure of the graph at the node. The subdivision of the region $R$ into cubic cells is the same as above, the key difference is in how to choose the surfaces to smear the triad filed. Instead of subdividing the boundary $\p C_I$ into cells $S^\alpha$, we now consider for each $C_I$ only three surfaces. We assign a local coordinate system $x^a$, and take the three surfaces to be any three surfaces $S_I^a$ inside the cube and orthogonal to each other.
The  Ashtekar and Lewandowski volume operator is defined as
\begin{equation}
\hat V (R) = \lim_{\eps\mapsto 0}\sum_{C_I} \sqrt{\left|\frac{1}{3!}\varepsilon_{ijk}\varepsilon_{abc} \hat E^i(S_I^a)\hat E^j(S_I^b)\hat E^k(S_I^c) \right|}.
\end{equation}
Unlike \Ref{ClassicalVolume}, this expression is already written in terms of fluxes, so we do not need the manipulation in terms of $W_I$.
We can directly use \Ref{fluxaction} to write
\begin{equation}
\label{VolAL}
\frac{1}{3!}\varepsilon_{ijk}\varepsilon_{abc} \hat E^i(S_I^a)\hat E^j(S_I^b)\hat E^k(S_I^c)= \frac{\hbar^3}{48}\varepsilon_{ijk}\varepsilon_{abc} \, \kappa_a\kappa_b\kappa_c \, J^i_{a}J^j_{b}J^k_{c},
\end{equation}
where again $J^i_{e_a}$ is the $SU(2)$ generator in the spin $j_a$ representation, and $\kappa_{S}$ is the relative orientation between the cell $S^a_I$ and the edge puncturing it, i.e. 
\begin{equation}
\kappa_{S}(e)= \left\lbrace \begin{array}{ll}
1 & \text{if }e\text{ lies above } S\\ 
-1 & \text{if }e\text{ lies below } S\\
0 & \text{otherwise}
\end{array} \right. 
\end{equation}
The operator \eqref{VolAL} strongly depends on the choice of the triple of surfaces $S_I^a$. In order to define a volume operator independent from the regulator, we average  \eqref{VolAL} on all the possible choices of orthogonal triples $S_I^a$ (see the \cite{Ashtekar:1997fb} for the details). 
The result is a sum over the nodes $n$ of the graph,
\begin{equation}\label{AL}
\hat V (R) = \hbar^{3/2} \sum_{n\in\Gamma} \sqrt{\left|\frac{\kappa_0}{48}\varepsilon_{ijk} \sum_{e,e',e''}\, \epsilon(e,e',e'')\, J^i(e)J^j(e')J^k(e'')\right|},
\end{equation}
where $e$,$e'$,$e''$ runs over the set of edges passing through the node $n$ and $\epsilon$ is the orientation function, which equals $0$ if the tangent directions of the three edges are linearly dependent and $\pm 1$ if they are linearly independent and oriented positively or negative respect to the orientation of $\Sigma$. $\kappa_0$ is an undetermined constant introduced by the averaging procedure.

This alternative volume differs from \Ref{V3} in two ways: firstly, it depends on an arbitrary constant, $\kappa_0$, unlike \Ref{V3} which is unique. Secondly, the absolute value is outside the internal summation, unlike \Ref{V3} where it is inside.
This second differences is what makes the AL operator sensible to the differential structure of the graph: in fact, the action of \Ref{AL} vanishes on nodes whose edges lie on a plane, because in that case $\epsilon\equiv 0$.

On the other hand, both operators annihilate a 3-valent node, and coincide (up to a proportionality constant) in the 4-valent case, where they are both proportional to $\varepsilon_{ijk} J^1_i  J^2_j J^3_k.$

\medskip

Summarizing, both volume operators act only on nodes of the graph. Their matrix elements vanish between different intertwiner spaces, and since every intertwiner space is finite dimensional, their spectra are discrete with minimal excitations proportional to the Planck length cube $\ell_P^3$.

Together with the discreteness of the area operator, these results show that in loop quantum gravity the \emph{space geometry is discrete at the Plack scale.} Each spin network describes a \emph{quantum geometry}, where each face dual to a link has an area proportional to the spin $j_e$, and each region around a node has a volume determined by the intertwiner $i_n$ as well as the spins of the link sharing the node.

It is important to stress that this is not a built-in discretization, as in lattice approaches to quantum gravity. It is a result of the quantum theory, similar to the quantization of the energy levels of an harmonic oscillator, or the radii of the atomic orbitals.

Thanks to this fundamental discreteness, where the minimal geometric excitations are proportional to the Planck length, the theory is expected not to have ultraviolet divergences, and to resolve the problem of the classical singularities of general relativity at the big bang (e.g. \cite{Cosmo}), or at the center of black holes (e.g. \cite{BH}). Addressing these issues of course requires taking into account the dynamics of the theory, to which we turn next.

\section{Loop quantum gravity: Dynamics}

\subsection{Solutions of the diffeomorphisms constraint}

Spin network states $\psi_{(\Gamma,j_e,i_n)}[A]$ are in $\HH^0_{kin}$, i.e. $\hat G^i \psi_{(\Gamma,j_e,i_n)}[A]=0$.
The next step in the Dirac program \Ref{Dirac1} is to implement the spatial diffeomorphisms, namely to find gauge-invariant states such that $\hat H^a \psi[A]=0$. 

To that end, consider a finite diffeomorphism $\phi$. Its action on the holonomy, as in \Ref{holdiffeoaction}, naturally induces an operator $\hat \phi$ on the space of cylindrical functions. This operator maps $Cyl_{\Gamma}$ to $Cyl_{\phi \circ \Gamma}$, that is $\hat\phi\psi_\Gamma=\psi_{\phi\circ\Gamma}$. Its action is well-defined and unitary, thanks to the fact that the Ashtekar-Lewandowski measure is diffeomorphism invariant. On the other hand, $Cyl_{\Gamma}$ and $Cyl_{\phi \circ \Gamma}$ are orthogonal Hilbert spaces, regardless of what the diffeomorphism is. This means that we can not define the action of an ``infinitesimal'' diffeomorphism
diffeomorphisms are all finite, from the perspective of cylindrical functions.
Although this might appear as a restriction, it is not a real problem for the construction of $\HH_{Diff}$. We can proceed by group averaging as we did for the Gauss constraint, and construct $\HH_{Diff}$ from those states invariant under \emph{finite} diffeomorphisms.

However, there are a couple of subtle issues to take into account. The first one has to do with the existence of symmetries of the graphs. Namely, for each graph there are always some diffeomorphisms that act trivially on it, leaving it unchanged. Let us distinguish two cases: the diffeomorphisms that exchange the links among themselves without changing $\Gamma$, call them $GS_\Gamma$ following \cite{AshtekarReview}, and those that also preserve each link, and merely shuffle the points inside the link,  call them TDiff$_\Gamma$. The latter have to be taken out, because their infinite-dimensional trivial action would spoil the group averaging procedure.

The second issue is that unlike imposing the Gauss law, imposing the invariance under diffeomorphism will not result in a subspace of $\HH^0_{kin}$, since diffeomorphisms are a non-compact group. Think for instance of $\psi(x)\in L_2[{\mathbbm R}, {\rm d}x]$ required to be invariant under translations (a non-compact group): the result is a constant function $c$, which is not in $L_2[{\mathbbm R}, {\rm d}x]$ since it is not integrable. It defines however a linear functional $c : \psi \in L_2[{\mathbbm R}, {\rm d}x] \mapsto {\mathbb C}$, since $\int {\rm d}x \, c\, \psi(x) = c\, \tilde \psi(0)$, the Fourier transform of $\psi$ evaluated in zero. Similarly, solutions to the diffeomorphism constraint can be described in terms of \emph{linear functionals} on $\HH^0_{kin}$. Let us denote $ \HH^{0*}_{kin}$ the space of linear functionals on  $\HH^0_{kin}$. Then, $\eta\in \HH^{0*}_{kin}$ is a diffeomorphic-invariant functional if
\be\label{diffinv}
\eta(\hat\phi \psi) = \eta(\psi) \quad \forall \psi \in \HH^0_{kin}.
\ee
The space of such functionals is denoted $\HH_{Diff}^*$, and the Hilbert space $\HH_{Diff}$ solution of the constraint is constructed by duality.
The condition \Ref{diffinv} should remind the reader of the analogue condition of gauge-invariance studied in the previous section. There, we were able to implement it via a simple group averaging procedure. In the same manner, we would like to define a projector ${\cal P}_{Diff}$ on $\HH_{Diff}$ such that
\be
\langle \psi | \psi' \rangle_{Diff} \equiv \langle \psi | {\cal P}_{Diff} | \psi' \rangle = 
\sum_{\phi\in{\rm Diff}/{\rm TDiff}_{\Gamma} } \langle\hat\phi\psi|\psi'\rangle,
\ee
where the sum is over all the diffeomorphism mapping $\Gamma$ into $\Gamma'$ \emph{except} those corresponding to the trivial ones TDiff$_\Gamma$.
See \cite{AshtekarReview} for more details.

\begin{figure}[ht]
  \begin{minipage}[t]{0.5\textwidth}
    \vspace{0.5cm}\includegraphics[width=5cm]{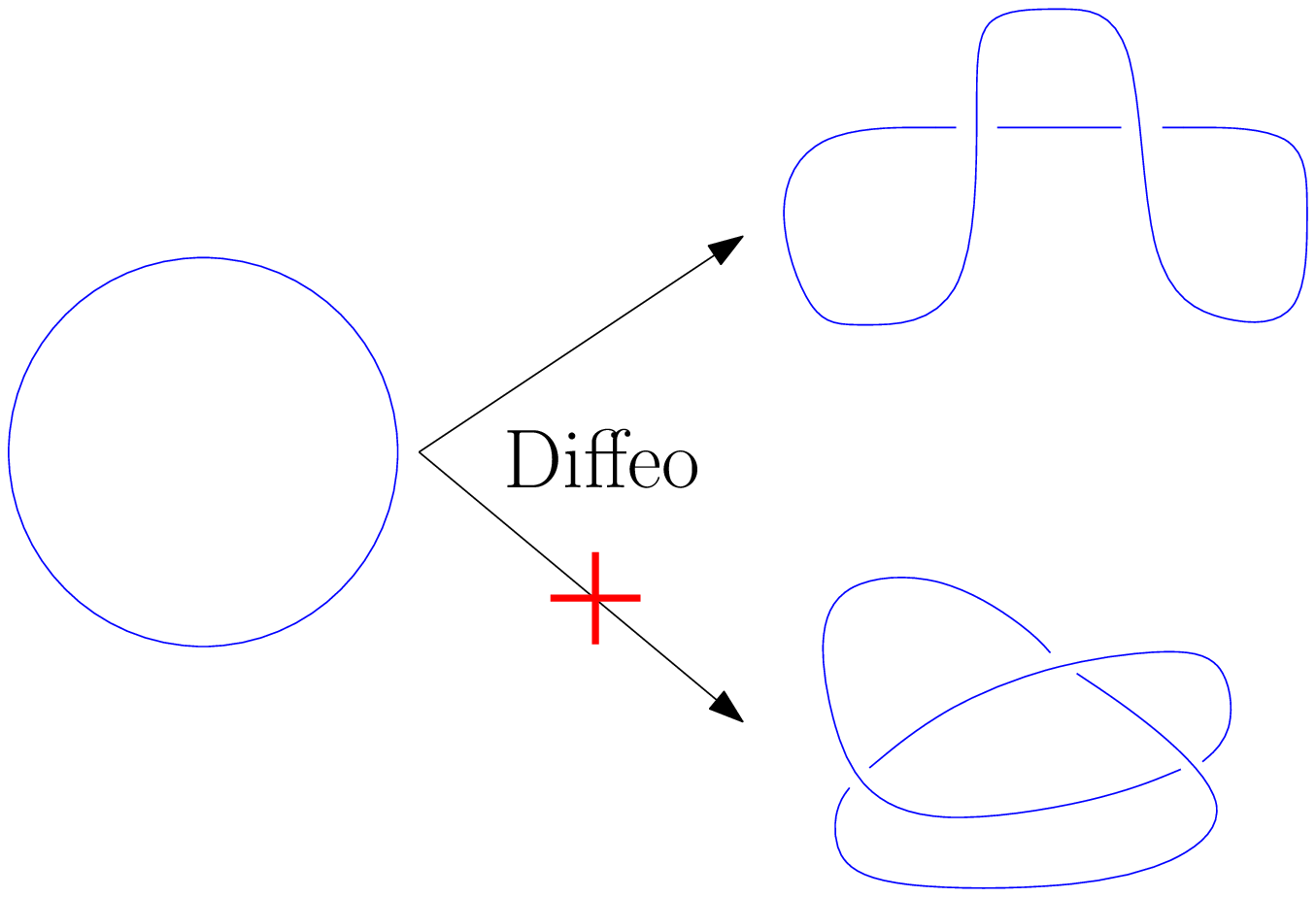}
  \end{minipage}
  \hfill
  \begin{minipage}[t]{0.5\textwidth}
    \vspace{0pt}\includegraphics[width=6cm]{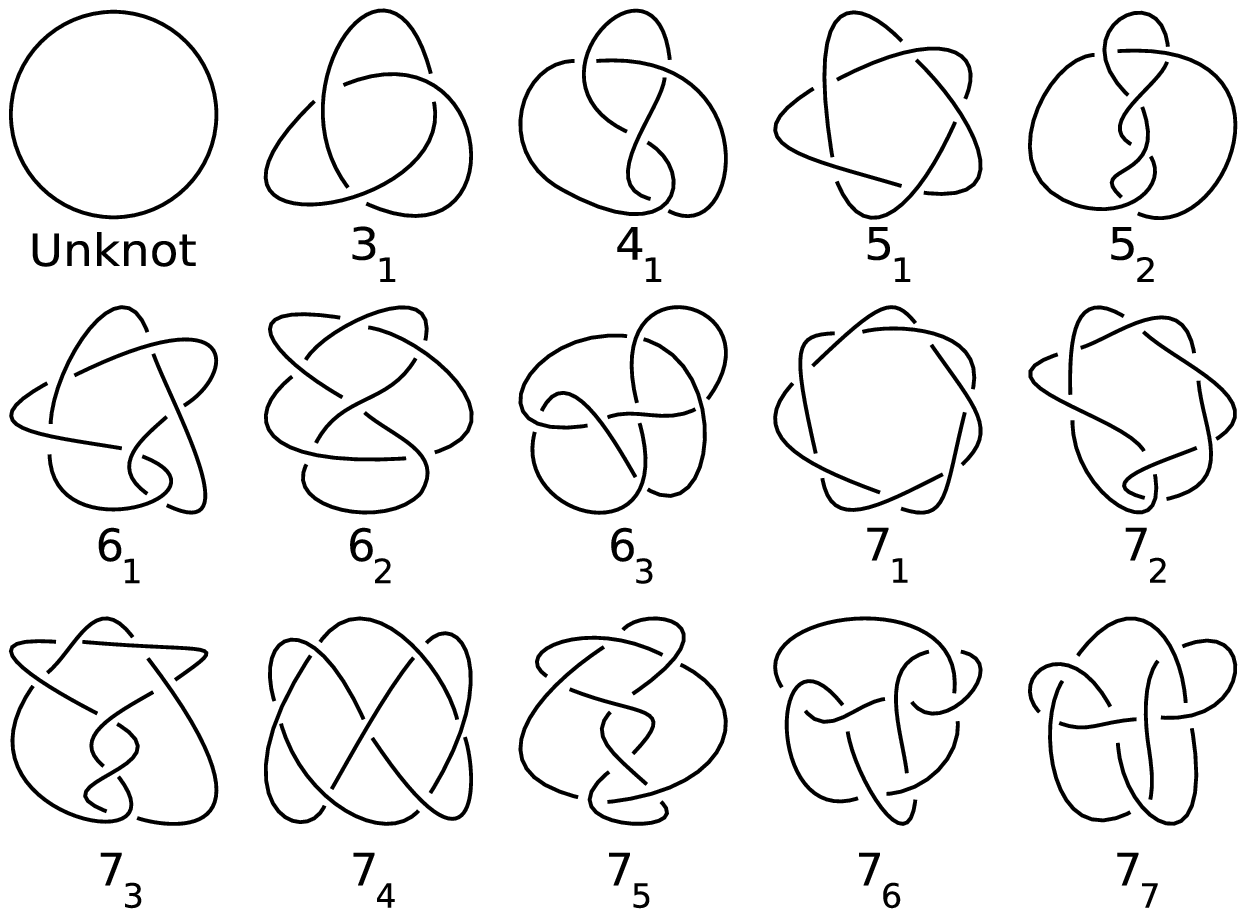}
  \end{minipage}
\caption{\label{Figknots} \emph{Left panel}. A diffeomorphism can change the way a graph is embedded in $\Sigma$, but not the presence of knots within the graph. \emph{Right panel}. The first few knots (without nodes), taken from Wikipedia. }
\end{figure}

The result of this procedure are spin network states defined on \emph{equivalence classes of graphs under diffeomorphisms}. These equivalence classes are called \emph{knots}, see Figure \ref{Figknots}. The study of knots forms an elegant branch of mathematics. The diff-invariant Hilbert space of loop quantum gravity is spanned by \emph{knotted spin networks}.

\subsection{The Hamiltonian constraint}

Finally, we approach the last step of Dirac's program. On the space of knotted spin networks $\HH_{Diff}$, we want to define the Hamiltonian constraint, and study its solutions.
The classical scalar constraint is
\begin{eqnarray}
{H} (N) &=& \int \mathrm{d}^3x N \epsilon^{ij}_k \frac{E^a_iE^b_j}{\sqrt{\mathrm{det}(E)}} \left(F^k_{ab} - 2 \left(1+\gamma^2\right) K^i_{[a}K^j_{b]}\right)  \nonumber \\
&=& {H}^E (N) - 2 \left(1+\gamma^2\right)  T(N),\label{LQGHam}
\end{eqnarray}
where we introduced the shorthand notation ${H}^E (N)$ and $ T(N)$. 
As with the ADM Hamiltonian constraint \Ref{ADMHam}, this expression is non-linear, which anticipates difficulties to turn it into an operator. However, a trick due to Thiemann \cite{ThiemannHam} allows us to rewrite \Ref{LQGHam} in a way amenable to quantization. Denoting $V=\int \sqrt{\mathrm{det}(E)}$ the volume of $\Sigma$, and
\begin{equation}
\bar{K} = \int K^i_a E^a_i,
\end{equation} 
we can use the classical brackets \Ref{AEalgebra} to establish the following identities,
\begin{eqnarray}
K^i_a &=& \frac{1}{\gamma}\left(A^i_a -\Gamma^i_a(E)\right) = \frac{1}{\gamma}\left\lbrace A^i_a,\bar K\right\rbrace, \\
\bar K &=& \frac{1}{\gamma^{3/2}}\left\lbrace \mathcal{H}^E(N\equiv 1),V\right\rbrace, \\
\frac{E^a_iE^b_j}{\sqrt{\mathrm{det}(E)}}\epsilon^{ijk}\epsilon_{abc} &=& \frac{4}{\gamma}\left\lbrace A^k_a,V\right\rbrace.
\end{eqnarray}
Using these relations, 
\begin{eqnarray}
{H}^E(N) &=& \int \mathrm{d}^3 x\; N \epsilon^{abc}\delta_{ij}F^i_{ab}\left\lbrace A^j_c,V\right\rbrace,\label{HE} \\
{T}(N) &=& \int \mathrm{d}^3 x\; \frac{N}{\gamma^3} \epsilon^{abc}\epsilon_{ijk} \left\lbrace A^i_a,\left\lbrace {H}^E(1),V\right\rbrace \right\rbrace \left\lbrace A^j_b,\left\lbrace H^E(1),V\right\rbrace \right\rbrace \left\lbrace A^k_c,V\right\rbrace. \nonumber
\end{eqnarray}

The advantage of this reformulation is that the non-linearity is mapped into Poisson brackets. The next step is to rewrite these expressions in terms of holonomies and fluxes, so that we can turn them into operators. Notice that we already know the volume operator, and its spectrum can be explicitly computed. This is very promising towards the prospect of knowing the action of the Hamiltonian constraint. Next, the connection and curvature have to be written in terms of holonomies. This requires a regularization procedure.
We describe it only for the first term, $H^E$, and refer the reader to the literature \cite{ThiemannBook} for $T(N)$. 

The connection can be easily expressed in terms of holonomies. Writing explicitly \Ref{holseries}, we have that for a path $ e_a$ of length $\epsilon$ along the $x^a$ coordinate, $h_{e_a}[A]\simeq 1 + \eps A^i_a \tau_i + O(\epsilon^2)$, therefore we also have
\begin{equation}\label{Adiscr}
h^{-1}_{e_a}\left\lbrace h_{e_a},V\right\rbrace  = \epsilon \left\lbrace A^i_a,V \right\rbrace + O(\epsilon^2).
\end{equation}
For the curvature, consider an infinitesimal triangular loop $\alpha_{ab}$, lying on the plane $ab$, and with coordinate area $\epsilon^2$. At lowest order in $\epsilon$,
\begin{equation}
h_{\alpha_{ab}} = 1 + \f12\epsilon^2 F^i_{ab} \tau^i + O(\epsilon^4),
\end{equation}
thus
\begin{equation}
h_{\alpha_{ab}} - h_{\alpha_{ab}}^{-1}=  \epsilon^2 F^i_{ab} \tau^i + O(\epsilon^4).\label{Fdiscr}
\end{equation}
At this point, we proceed as we did for the geometric operators in the previous section. We introduce a cellular decomposition of $\Sigma$, and regularize the integral as a Riemann sum over the cells $C_I$, 
\begin{align}
\label{regClassicalHC}
{H}^E =& \lim_{\epsilon \to 0} \sum_I \epsilon^ N_I3 \epsilon^{abc} \mathrm{Tr}\left(F_{ab}\left\lbrace A_c,V\right\rbrace  \right) =\\
=& \lim_{\epsilon \to 0} \sum_I N_I \epsilon^{abc} \mathrm{Tr}\left(\left( h_{\alpha_{ab}} - h_{\alpha_{ab}}^{-1}\right) h^{-1}_{e_c}\left\lbrace h_{e_c},V\right\rbrace\right). 
\end{align}
This time is more convenient to specify the cellular decomposition in terms of a \emph{triangulation}, namely a collection of tetrahedral cells. The loop $\alpha_{ab}$ can then be adapted to the triangular faces of this decomposition, as in the following figure.
\begin{center}
\includegraphics[scale=1]{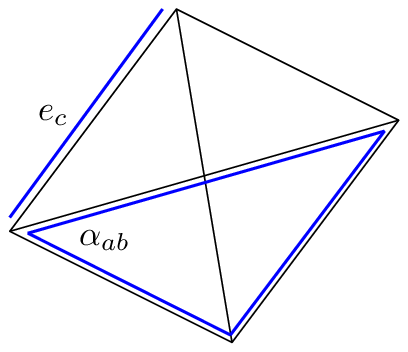} 
\end{center}

This expression can now be promoted to an operator in the quantum theory, 
\begin{equation}
\label{regQuantumHC}
\widehat{{H}}^E 
= \lim_{\epsilon \to 0}\sum_I N_I \epsilon^{abc} \mathrm{Tr}\left(\left( \hat h_{\alpha_{ab}} - \hat h_{\alpha_{ab}}^{-1}\right) \hat h^{-1}_{e_c}\left[ \hat h_{e_c},\hat V\right]\right).
\end{equation}
This is a well-defined operator, whose action is explicitly known. 
It inherits the property of the volume operator of acting only on the nodes of the spin network. From the holonomies, it modifies the spin network by creating new links carrying spin 1/2 around the node, see the following figure. 
 \begin{center}
\includegraphics[scale=1]{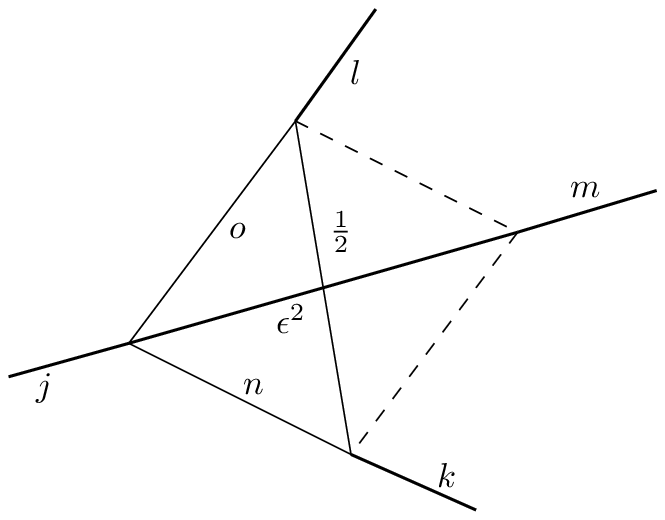} 
\end{center}
Finally, its amplitude depends on the details of the action of the volume operator. See the literature \cite{RovelliBook,ThiemannBook,AshtekarReview} for details.

Since the only dependence on $\epsilon$ is on the position of the new link, in the Hilbert space $\mathcal H_{Diff}$ where the position of the link is irrelevant, the limit can be safely taken without affecting the result.
Not only this operator is well-defined, also the Dirac algebra \Ref{ConstraintAlgebra} can be realized at the quantum level. To see this, we need first to note that the new ``exceptional'' links added by \Ref{regQuantumHC} carry zero volume and are thus 
invisible to a further action by \Ref{regQuantumHC}. This crucial property requires the Ashtekar-Lewandowski version of the volume quantum operator, because the new nodes at the junction of the new link with the old ones are planar, and the AL volume operator vanishes on planar nodes. Thanks to this, $\widehat{\mathcal{H}}(N_1)$ and $\widehat{\mathcal{H}}(N_2)$ commute on the space $\HH_{Diff}$ of diffeomorphic-invariant spin networks. Hence for diffeomorphism invariant states $\left\langle \phi \right|\in \mathcal{H}_{Diff}$,
\begin{equation}
\left\langle \phi \right| \left[\widehat{\mathcal{H}}(N_1),\widehat{\mathcal{H}}(N_2) \right] \left| \psi \right\rangle = 
0 = \left\langle \phi \right| \mathcal{H}(N_{[1}\vec \nabla N_{2]}) \left| \psi \right\rangle, \qquad \forall \psi \in \mathcal{H}_{kin}.
\end{equation}
This is the correct commutator algebra on-shell. Thanks to this non-trivial property, the quantization is said to be anomaly-free.

Another key result is that an infinite number of states solutions of $\widehat{{H}}$ are known: any graph without nodes is in the kernel of $\widehat{{H}}^E$ and $\widehat{{T}}$. This result \cite{RS1}, together with the discreteness of the spectra \cite{RS2}, started the whole interest in the loop quantum gravity approach.
More in general, a formal solution of the Hamiltonian constraint will be a particular linear combination (generally infinite) of spin networks with an arbitrary number of exceptional links, whose coefficients depend on the details of the quantum operator and on the spins carried by the spin network,
\begin{center}
\includegraphics[scale=0.6]{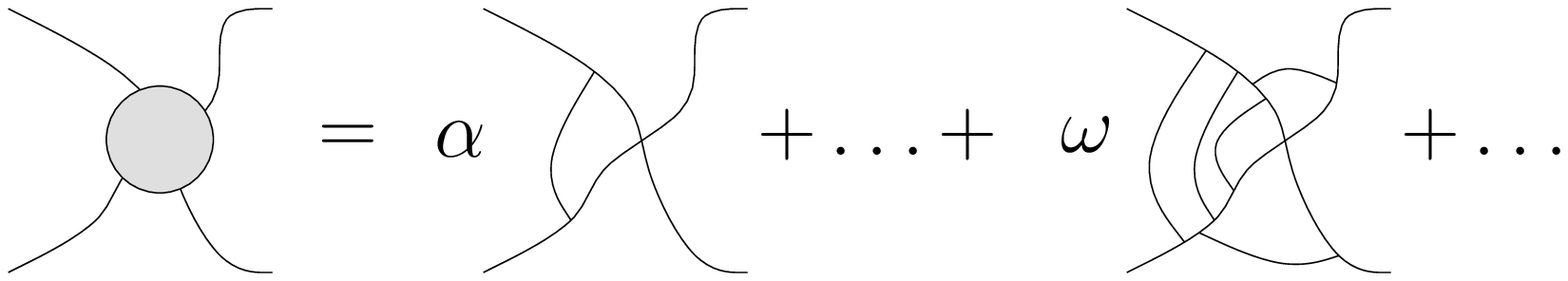} 
\end{center}
This construction defines a new spin network with a ``dressed node'' as the solution. However, this procedure is only formal, and no explicit solutions are known in general.

In conclusion, we have a perfectly well-defined Hamiltonian constraint, whose action is explicitly known and \emph{finite}. 
An infinite number of states solutions of $\widehat{{H}}$ are known, and the Dirac algebra is anomaly-free on physical states. 
Compare this with the old-fashioned Wheeler-De Witt equation \Ref{WdW}, which was badly ill-defined, and we see the full force of the use of the Ashtekar variables to quantize general relativity. 

Nevertheless, the program of quantization is far from being complete: the complete characterization of $\HH_{phys}$ nor the full spectrum of $\widehat{H}$ are known. In spite of the successes of this approach, some limitations need to be stressed. In particular, although well-defined, the Hamiltonian constraint is plagued by a number of ambiguities:
\begin{itemize}
\item We can change the spin of the exceptional edges to an arbitrary value $j$. This change have significant physical effects in loop quantum cosmology for example.
\item We can regularize the connection and the curvature with different paths and still make a consistent theory \cite{AshtekarReview}. In particular, one can envisage a different construction in which the constraint acts on more than one node simultaneously, as advocated in \cite{SmolinHam}.
\item Alternative orderings can be explored.
\end{itemize}

\subsection{Current approaches}

The effort to gain control over these ambiguities has flourished into two main lines of research. The first one is the idea of the Master Constraint. There one defines a unique constraint implementing simultaneously the diffeomorphisms and scalar constraints. The second one is the spin foam formalism. There one abandons the canonical approach, and seeks a functional integral description of transition amplitudes between spin network states. These two approaches have seen recent important developments, and are the current state of the art of the field. We will unfortunately not have time to review them here, but the interested reader can consider references \cite{ThiemannBook, ThiemannMC} for the first approach, and \cite{RovelliNL} for the second.

To conclude, although the kinematics of loop quantum gravity is beautifully under control, the dynamics is still work in progress. 
We remark nonetheless that the key result on space discreteness, derived using kinematical states, is expected to hold also for physical states \cite{RovelliBook}. In the next and final Section, we want to go back to this discreteness and make some comments about it.

\section{On quantum geometry}

Let us review what we have learned in Section \ref{SecKin}. The golden tools of loop quantum gravity are the spin network states. These form a basis in the kinematical Hilbert space, and diagonalize geometric operators. In particular, the quantum numbers carried by a spin network, $(\Gamma,j_e,i_n)$, define a notion of \emph{quantum geometry},\footnote{Each face dual to a link has an area proportional to the spin $j_e$, and each region around a node has a volume determined by the intertwiner $i_n$ as well as the spins of the link sharing the node.} in the same way as the quantum number of an harmonic oscillator defines its quantum state.
These quantum numbers can be compared with the kinematical metric $g_{ab}$, defining the classical geometry of space. We use the word kinematical to mean that $g_{ab}$ is an arbitrary metric, not necessarily a solution of Einstein's equations, just like an arbitrary spin network spans the kinematical Hilbert space, not necessarily solving the diffeomorphisms and Hamiltonian constraints.
The quantum geometry described by $(\Gamma,j_e,i_n)$ is very different than a classical geometry $g_{ab}$.
Specifically, it is largely insufficient to reconstruct a metric $g_{ab}$. 
We can highlight three key differences:
\begin{enumerate}
\item[(i)] Quantized spectra: the spectra of geometric operators are discrete, as opposed to the continuum values of their classical counterparts.
This is a standard situation in quantum mechanics, think for instance of the discretization of the energy levels of an harmonic oscillator.
\item[(ii)] Non-commutativity: not all geometric operators commute among themselves. This is a consequence of the non-commutativity of the fluxes \Ref{Enc}. This is also standard, like the incompatibility of position and momentum observables.
\item[(iii)] Distributional nature: the states capture only a finite number of components of the original fields, that is their values along paths (for the connection) and surfaces (for the triad). This is reminiscent of what happens in lattice theories, where the continuum field theory is discretized on a fixed lattice and only a finite number of degrees of freedom are captured. 
\end{enumerate}

In spite of these differences, if the theory is correct it must admit a \emph{semiclassical} regime where a smooth geometry emerges. And furthermore, the dynamics of this smooth geometry should be given by general relativity, at least in some  approximation. This is what we expect from any quantum theory: that the $\hbar\mapsto 0$ limit is well-defined, and reproduces classical physics. The points (i-iii) show that this limit might not be trivial to obtain in loop quantum gravity. Although we started from genuine general relativity and quantum mechanics, without any exotic ingredient, we ended up with a description in terms of quantities, $(\Gamma,j_e,i_n)$, which are very far from the original ones, $g_{ab}$.
This is the problem of the semiclassical limit.

In this final Section, we would like to comment on the first part of this problem, namely the emergence of a smooth geometry. This requires understanding the points (i-iii). The first point is easier to deal with: also the orbitals of the hydrogen atoms are quantized, placed at distances labeled by an integer $n$. The classical Keplerian behavior is recovered if we look at the large $n$ limit. Similarly, continuum spectra are recovered in the large spin limit $j_l\mapsto \infty$.
Points (ii) and (iii) are more subtle. The key to deal with them is the use of \emph{coherent states}, namely linear superpositions of spin network states peaked on a smooth geometry. 

Recall that coherent states are peaked on a point in the classical phase space. For instance, the coherent state $|z\rangle$ for an harmonic oscillator is peaked on the initial position $q={\rm Re}(z)$ and momentum $p={\rm Im}(z)$. The phase space is the familiar cotangent bundle to the real line, $(q,p) \in T^* R$. Analogously, coherent states for loop quantum gravity are peaked on a point $(A^i_a(x), E^a_i(x))$ in the classical phase space, which defines an intrinsic (through the triad) and extrinsic (through the connection) 3-geometry. Such coherent states for loop quantum gravity were introduced in \cite{CS}. These states minimize the uncertainty of the flux operators, thus addressing (ii). In order to address (iii) and recover a smooth geometry \emph{everywhere} on $\Sigma$, the coherent states have support over an infinite number of graphs.

On the other hand, dealing with an infinite number of graphs is a formidable task, and for practical purposes, one often needs to rely on approximations. A convenient one is to allow only states living on a fixed finite graph $\Gamma$. The Hilbert space ${\cal H}_\Gamma$ provides a \emph{truncation} of the theory, which may be sufficient to capture the physics of appropriate regimes. Within this truncation, one considers coherent  states in ${\cal H}_\Gamma$. However, in  which sense one can assign a classical geometrical interpretation to these states?
They will be peaked on points in the phase space corresponding to $\HH^0_\Gamma$, which consists of classical holonomies and fluxes on $\Gamma$. These quantities capture only a \emph{finite} number of components of a continuum geometry. 
To be able to interpret these truncated coherent states in a physical sense, we need to use these data to approximate a continuum geometry.
The problem is similar to approximating a continuous function $f(x)$ if we are given a finite number of its values, $f_n=f(x_n)$.
Various interpolating procedures are common. We now describe a notion of interpolating geometry which emerges naturally.

The associated gauge-invariant Hilbert space on a fixed graph is a sum over 
intertwiner spaces $\HH_n \equiv {\rm Inv}[\underset{e \in n}{\otimes} V^{(j_e)}]$ associated to the nodes (see \Ref{Hkin0}),
\be\label{Hs}
{\cal H}^0_\Gamma = \oplus_{j_l} \left(\otimes_n {\cal H}_{n} \right).
\ee 
Just like $L_2[R, {\rm d}x]$ is the quantization of the classical phase space $T^*R$, this Hilbert space is the quantization of a classical phase space\footnote{Up to singular points, but this is not important for the following.}, denote it  
$S_\Gamma$. 
The important result which is relevant for us is that this phase space can be parametrized as follows \cite{Twi},
\be\label{S}
S_\Gamma = \underset{l}\times \, T^*S^1 \underset{n}\times {\cal S}_{F(n)}.
\ee
Here $T^*S^1$ is the cotangent bundle to a circle, $F$ the valency of the node $n$, and ${\cal S}_F$ is a $2(F-3)$-dimensional phase space, introduced by Kapovich and Millson \cite{Kapovich}.
This parametrization mimics the decomposition \Ref{Hs}. In particular, the quantization of $T^*S^1$ gives the quantum number $j_l$ associated to each link, and the quantization of ${\cal S}_{F(n)}$ gives the $F-3$ intertwiners associated to an $F$-valent node.

\begin{figure}[h!]
\begin{center}\includegraphics[scale=0.4]{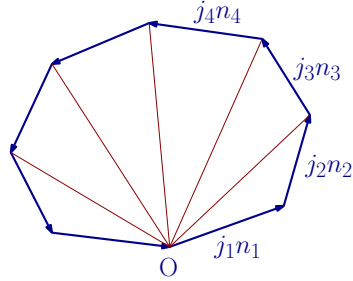}\end{center}
\caption{A polygon in ${\mathbb R}^3$, with $F$ sides of fixed lengths. Because of the closure condition, not all the components of the normals are independent, but only $2F-3$. If we further consider the polygon up to rotations, that is the space ${\cal S}_F$, this is only specified by $2F - 3 - 3 = 2(F-3)$ numbers. These can be taken to be the lengths of $F-3$ diagonals and their conjugated dihedral angles (which measure the way the polygon hinges on each diagonal).\label{FigPolygon}} 
\end{figure}

The parametrization defines a notion of interpolating geometry associated to a cellular decomposition of $\Sigma$ dual to $\Gamma$, called \emph{twisted geometry} \cite{Twi}. The key to the geometric interpretation of \Ref{S} is the node space ${\cal S}_{F(n)}$. Fix the spins $j_l$ on the links connected to the node, and consider $F$ unit vectors $n_l$ in ${\mathbb R}^3$ constrained to satisfy the following \emph{closure condition},
\be\label{closure}
C = \sum_{l=1}^F j_l n_l = 0.
\ee
Because of the closure, the vectors $j_l n_l$ span a polygon in ${\mathbb R}^3$, see Figure \ref{FigPolygon}.
The space of all polygons with fixed $j_l$ up to SO(3) rotations forms the phase space ${\cal S}_F$, called by Kapovich and Millson the space of shapes of the polygon.

This means that we can interpret the intertwiners as ``quantized polygons''. Such interpretation is appealing, but not particularly useful for our purposes. More promising would be a description in terms of \emph{polyhedra}, rather than polygons: Polyhedra can be glued together to approximate a smooth manifold, and their geometry will induce a discrete metric of some sort. 

The good news \cite{noi} is that this is precisely what happens! In fact, each (non-coplanar) configuration of vectors $n_l$ describes also a \emph{unique polyhedron} in ${\mathbb R}^3$, with $j_l$ as the areas of the faces, and $n_l$ as their unit normals. More precisely, a convex bounded polyhedron.
The explicit reconstruction of the polyhedron geometry from holonomies and fluxes is studied in \cite{noi}.
\begin{figure}
\begin{center}\includegraphics[scale=0.2]{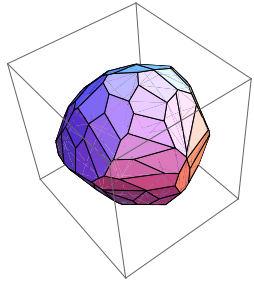}\end{center}
\caption{A generic polyhedron with 100 faces. Notice that most faces are hexagons.} 
\end{figure}
Therefore, up to the degenerate configurations corresponding to coplanar normals, we can visualize ${\cal S}_F$ as a polyhedron with $F$ faces.

Going back to \Ref{S}, the $T^*S^1$ space on the links carries information needed to define the parallel transport between adjacent polyhedra.
This unique correspondence between ${\cal S}_F$ in \Ref{S} and polyhedra means that \emph{each (non-degenerate) classical holonomy-flux configuration on a fixed graph can be visualized as a collection of adjacent polyhedra, each one with its own frame, and with a notion of parallel transport between any two polyhedra}. 
Since twisted geometries parametrize the classical phase space on a fixed graph, coherent states on a fixed graph are peaked on the approximated smooth geometry described by a twisted geometry, i.e. a collection of polyhedra. 

\begin{figure}[h!]
\begin{center}\includegraphics[scale=0.25]{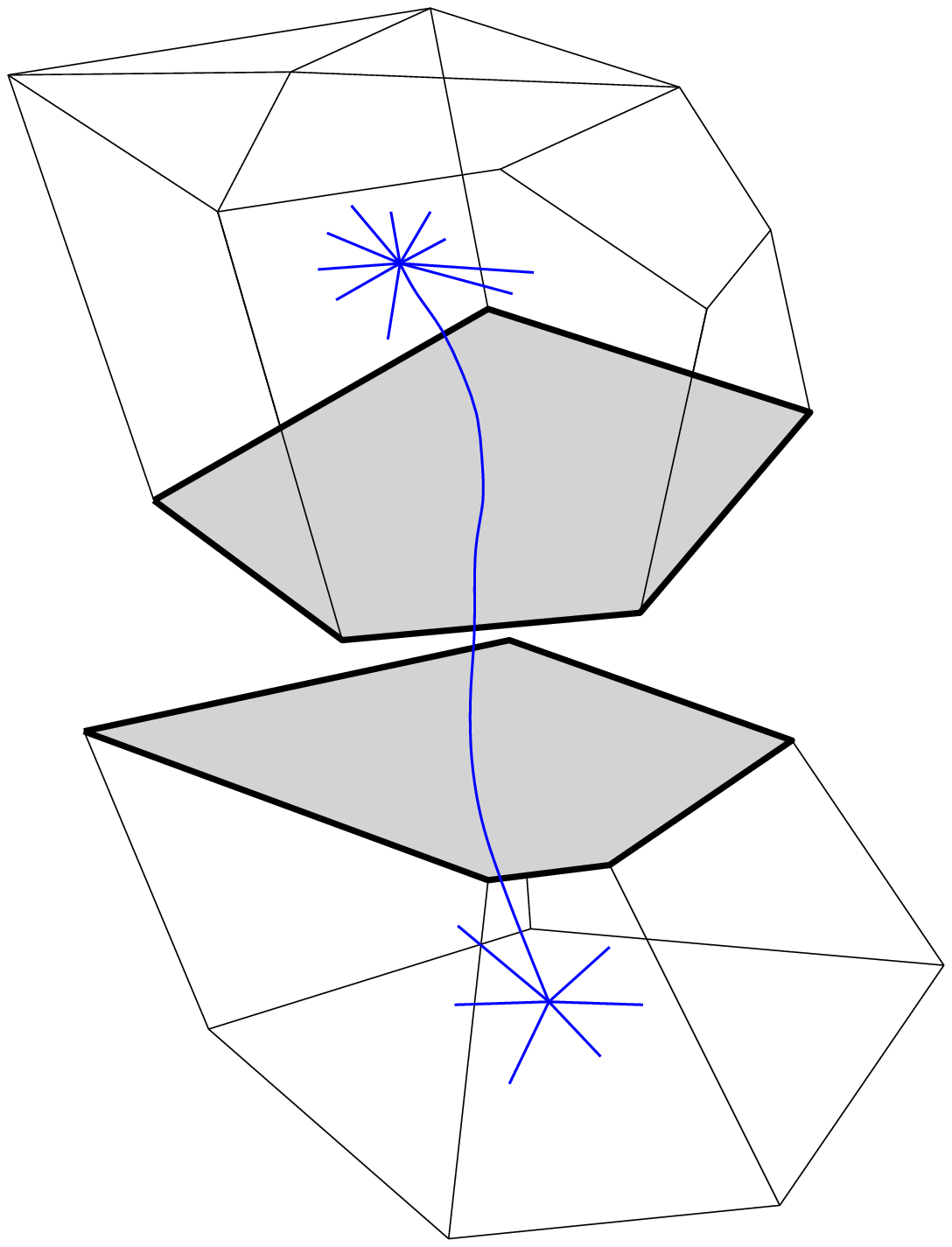} \hspace{2cm} \includegraphics[scale=0.3]{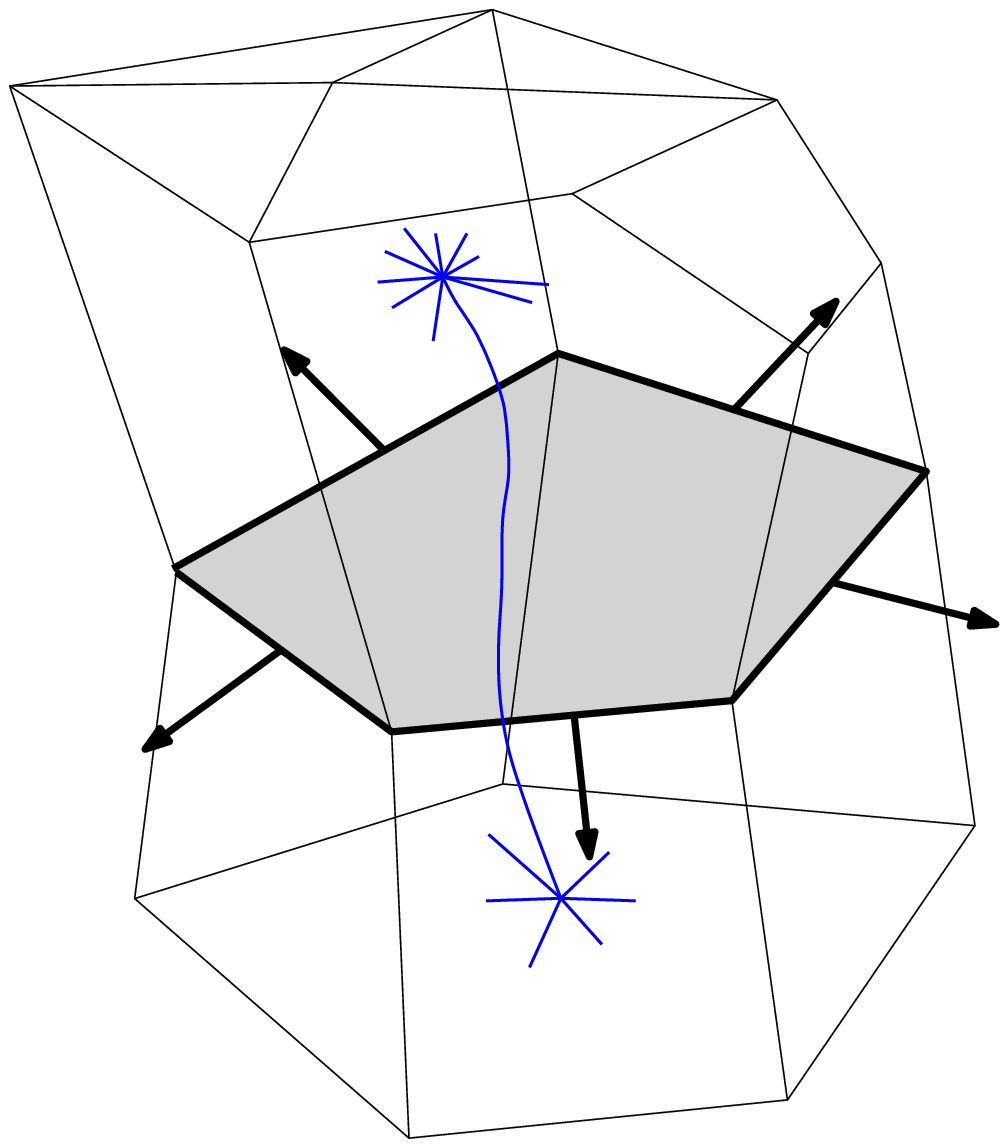}\end{center}
\caption{Two high-valency nodes and the link connecting them. Left panel: In general, the two adjacent polyhedra, defined by holonomies and fluxes on the nodes, don't glue well. Even if the area of a polygonal face is the same because there is a unique spin associated with the link, the shapes will be different in general. Right panel: in order for the shapes to match, one needs to impose appropriate conditions on the polyhedra such as the matching of the normals to the edge in the plane of the face. These conditions, studied in \cite{noi}, affect the global shape of the polyhedron.\label{Figaa}} 
\end{figure}

Twisted geometries have a peculiar characteristic which justifies their name: they define a metric which is locally flat, but \emph{discontinuous}. To understand this point, consider the link shared by two nodes. Its dual face has area proportional to $j_l$. However, its shape is determined independently by the data around each node (the normals and the other areas), thus generic configurations will give different shapes. In other words, the reconstruction of two polyhedra from holonomies and fluxes does not guarantee that the shapes of shared faces match. Hence, the metric of twisted geometries is discontinuous in the sense that the shape of a face ``jumps'' when going from one polyhedra to the next. See left panel of Figure \ref{Figaa}.

Notice that one can also consider a special set of configurations for which the shapes match, see right panel of Figure \ref{Figaa}. This is a subset of $S_\Gamma$ which corresponds to piecewise flat and continuous metrics. For the special case in which all the polyhedra are tetrahedra, this is the set-up of Regge calculus, and those holonomies and fluxes describe a 3d Regge geometry. The matching conditions in this case were studied in \cite{Dittrich}.
For an arbitrary graph, the matching conditions are studied in \cite{noi}, and the result would be Regge calculus on generic cellular decompositions.

The relation between twisted geometry and Regge calculus implies that holonomies and fluxes carry \emph{more} information than the phase space of Regge calculus. We stress that this is not in contradiction with the fact that the Regge variables and the LQG variables on a fixed graph both provide a truncation of general relativity: simply, they define two distinct truncations of the full theory. 
See \cite{IoCarlo} for a discussion of these aspects.

Summarizing, when we truncate the theory to a single graph, we capture only a finite number of degrees of freedom of the geometry. This finite amount of information can be suitably interpolated to give an approximate description of a smooth geometry.
The reparametrization in terms of twisted geometries achieves a prescription for this interpolation procedure: Starting from holonomies and fluxes on a graph, we can assign to them a specific twisted geometry, that is, a bunch of polyhedra stuck together.

These results offers on the one hand a new way to visualize the quantum geometry of loop quantum gravity in terms of fuzzy polyhedra.
On the other hand, they open the door to new beautiful mathematical ingredients, from the geometry of polyhedra \cite{Alexandrov, noi}, to twistors \cite{Twi2}, to new emerging symmetries \cite{Freidel}.

These intriguing aspects, together with the striking recent developments in the study of the dynamics and in the applications to cosmology and the physics of black holes, make of loop quantum gravity an excellent arena for a student interested in the problems of physics at the Planck scale.

\vspace{1cm}

{\bf Acknowledgements} SS wishes to thank the organizers of the School for their kind invitation, and for their splendid hospitality and great help in making the time in Algeria so enjoyable. J'espere que pour la prochaine ecole, nous aurons aussi le tournoi de baby-foot!


\end{document}